\documentclass{elsart}

\usepackage{epsfig}

\usepackage{amssymb}

\newcommand{\lsim}{\mathrel{\mathop{\kern 0pt \rlap
  {\raise.2ex\hbox{$<$}}}
  \lower.9ex\hbox{\kern-.190em $\sim$}}}
\newcommand{\gsim}{\mathrel{\mathop{\kern 0pt \rlap
  {\raise.2ex\hbox{$>$}}}
  \lower.9ex\hbox{\kern-.190em $\sim$}}}

\begin{document}

\begin{frontmatter}



\title{Liquid Noble gases for Dark Matter searches: a synoptic survey.}


\author{R. Bernabei$^{a,b}$\corauthref{*}}
\author{, P. Belli$^b$,}
\author{A. Incicchitti$^c$, D. Prosperi$^{d,c}$}
\address[a]{Dipartimento di Fisica, Universit\`a di Roma ``Tor Vergata'',I-00133 Rome, Italy}
\address[b]{INFN, Sezione di Roma ``Tor Vergata'', I-00133 Rome, Italy}
\address[c]{INFN, Sezione di Roma, I-00185 Rome, Italy}
\address[d]{Dip. di Fisica, Universit\`a di Roma ``La Sapienza'',  I-00185 Rome, Italy}

\corauth[*]{Corresponding author.{\it Email address:} rita.bernabei@roma2.infn.it (R. Bernabei).}

\begin{abstract}
A technical and metho\-do\-logical comparison of liquid noble gas experiments is presented
 and the low energy physics application of double phase noble gas detectors in direct Dark Matter investigations is discussed.

\end{abstract}

\end{frontmatter}

\section{Introduction}\label{Int}

Liquid noble gas detectors have been proposed and used in different fields
of research for many years \cite{ref1,Bar}. The use of liquid xenon
as a pure scintillator in direct Dark Matter particle investigations dates back to 1990 with the
pioneering work by DAMA/LXe  \cite{dam1} and later on by the UK coll. \cite{zep0}.
Other authors have also expressed an interest in using liquid argon or xenon, see e.g. ref. \cite{Cli1}.

There has recently been a great deal of interest in double phase noble gas detectors, i.e. detectors exploiting the primary
scintillation in the liquid phase and the secondary proportional scintillation in the gas phase, as their use might be possible in one approach to direct Dark Matter investigations, the detection of WIMP (Weakly Interacting Massive Particles) 
candidates \cite{GW} in the case of their interaction would  only produce recoil nuclei. 

Different kinds of WIMP
interaction have been studied: interactions with the target nuclei giving rise to elastic scattering (with spin independent 
and/or spin dependent coupling), inelastic scattering, preferred inelastic interaction (with the Dark 
Matter particle that goes into an excited state instead of the nucleus) \cite{SW},  the impact of a partial electromagnetic contribution 
due to ionisation and excitation of bound atomic electrons 
induced by the presence of the recoiling atomic nucleus (Migdal effect) \cite{Mig}, etc.. It has also been indicated that electromagnetic
 interaction with atomic electrons may occur, 
when WIMP interaction with the nucleus is inhibited \cite{ele}.

Many
other particles and types of interaction have been considered and Dark Matter could even be of a multicomponent nature.
Varying lines of research also suggest that different kinds of interaction require suitable levels of sensitivity for their detection. 
 For example a  large number of papers on axion-like particles have indicated that although they share similar phenomenology with ordinary matter like the axion,
they have significantly different mass and coupling constants. Hence in this case, detection is based on the total conversion of the absorbed bosonic mass
into electromagnetic radiation \cite{bos}. In the framework of warm dark matter, light dark matter candidates have also been considered \cite{ldm}, inelastic 
scattering channels on the electron or on the nucleus have been studied for axino, sterile neutrinos and even MeV-scale LSP in Susy theories. 
After inelastic interaction, a lighter particle is produced and the target recoil energy is the detectable quantity. The target can be an electron or a nucleus, so the detectable quantity is
due to a different mechanism and has different features.

On the other hand, experiments exploiting the dual phase noble gas detection considered in this paper,
such as those based on the bolometer + ionisation/scin\-til\-lation detection method,
only focus on WIMPs
and one type of WIMP interaction: the elastic scattering process on the target nuclei.
However, when the target only has isotopes with even nuclei, such 
as is the case for natural argon and neon (the latter is practically even as $^{21}$Ne is only 0.27\%) they are not
 sensitive to spin dependent WIMP-nucleus inte\-ra\-ctions at all\footnote{Even for nuclei sensitive to spin dependent interaction
 different sensitivities are expected among odd-nuclei having an 
unpaired proton (for example, $^{23}$Na and $^{127}$I) and odd-nuclei having an unpaired 
neutron (for example, the odd Xe and Te isotopes and $^{73}$Ge).
This is due to uncertainties concerning the choice of  the nuclear potential,  the effective WIMP-nucleon coupling strengths and form and spin factors.}.
Hence nuclear recoil is the only process that can be considered in these experiments
 and the only feature that might be detected.
As a result, an exclusion plot in the plane cross section versus WIMP mass is usually produced in a fixed single model framework 
of experimental and phenomenological parameters without taking into account any uncertainty or alternative choices, and assuming 
that ideal results have been obtained using the several applied procedures.
This is a significant limitation to any investigation that wishes to explore Dark Matter effectively.
Due to these and other arguments no model independent comparison can be drawn from results obtained using different
 target materials and/or different experimental approaches.
  
Three gases have been discussed with regard to two phase detectors:
xenon (XENON, ZEPLIN), argon (WARP) and neon (SIGN) and recently measurements have been
published for the ZEPLIN-II, ZEPLIN-III, XENON10 and WARP prototypes.  Although these state of art prototypes are at the R\&D stage, successors
 on ton scale are already being discussed (ArDM, ELIXIR, LUX).

This paper looks at the low energy experimental application 
of liquid noble gases and compares features of the apparata such as design and  performance,
purity and radiopurity, trigger, calibration, data reduction, rejections and analyses.
The data  published up to now and examined here  correspond to an
 exposure of 136 kg $\times$ day for XENON10, 225 kg $\times$ day for ZEPLIN-II, 847 kg $\times$ day for ZEPLIN-III and 96.5  kg $\times$ day for WARP.
This paper compares the performance of apparata as published by the authors or presented at international 
conferences. It is worth underlining that technical aspects are extremely important as they can affect experimental 
results and should be studied accurately before high experimental sensitivity is claimed.\\
It is hoped that the discussion on double phase detectors will prove useful for commissioned  projects, projects under construction and future projects.

In order to avoid any misunderstanding, any published sentences quoted in this paper are written in italics.

\section{Experimental techniques}\label{Dat}

High electron mobility in liquid noble gases was discovered in 1948 \cite{Dav} while 
the first studies of their scintillation properties  dates back to the 1950s \cite{Nor}
when the scintillation properties and relative efficiencies of gas, solid and liquid xenon, krypton and argon were considered with respect to NaI(Tl) 
and liquid xenon emerged as a particularly promising candidate \cite{Bar,Dok4,noi2}. 
There is a great deal of literature concerning the scintillation properties of noble gases and we recall a few of interest.
 The scintillation light of liquid noble gases is emitted at 178 nm for liquid xenon, 128 nm for liquid argon and  77 nm for liquid neon\cite{Doksci,Pack}.
The mean energy to produce a scintillation photon in the case of $\alpha$ particle excitation is W$_s$ = 27.1 eV (liquid Ar) and 17.9 eV (liquid Xe),
 while in the case of electron excitation W$_s$ = 24.4 eV (liquid Ar) and 21.6 eV (liquid Xe) \cite{Doksci}. 
 Less data are available on  liquid neon: $\alpha$  particle W$_s$ is 169 eV and the electron W$_s$ is 135 eV \cite{Mich}.

In the years that followed, electron mobility was studied in more detail \cite{Mar,Alv,Dok2}, and many studies focussed on ionisation;
 the mean energy to produce an ion-electron pair is W= 19.5 eV (liquid Ar) and W= 14.7 eV (liquid Xe) \cite{Doksci}.
 Moreover electron multiplication was discovered \cite{Der} and it was shown that ``proportional scintillation'' 
 is present in the liquid phase as well as in the gaseous one \cite{Lan}. This led to the testing of both direct and proportional scintillation
 in liquid xenon by several groups including XelTpc which was part of the I.N.F.N. V technology committee \cite{Mas,coi,ica}.

Anti-correlation between scintillation and ionisation signals was observed in liquid xenon and argon \cite{Ku78,Dok3}.
Anti-correlation occurs when 
ionisation increases and scintillation decreases or vice versa
and the combined signal from these two processes is proportional to the
deposited energy.
 Different patterns are present for relativistic heavy particles and 1 MeV electrons\cite{Ku78,Dok3}. In the latter case, compensation between ionisation and scintillation 
only occurs in high electric fields. In general, scintillation and ionisation are  
 complementary and the effect of an external field is to suppress the recombination process,
 with the result of producing more free electrons and less scintillation photons. 
Experimental work and configurations have shown that the energy 
resolutions obtained for xenon and argon are worse than those by theoretical estimated limits  \cite{Dok1}. 
The reasons for this are not yet
fully understood, and many aspects are still under discussion \cite{Apr04}.
In fact despite progress in this field, it is still not completely possible to simultaneously measure scintillation and ionisation signals; 
difficulties arise as a result of electronic noise in charge readout and UV light collection.

\begin{figure}
\begin{center}
\includegraphics[height=11.0cm,angle=-90]{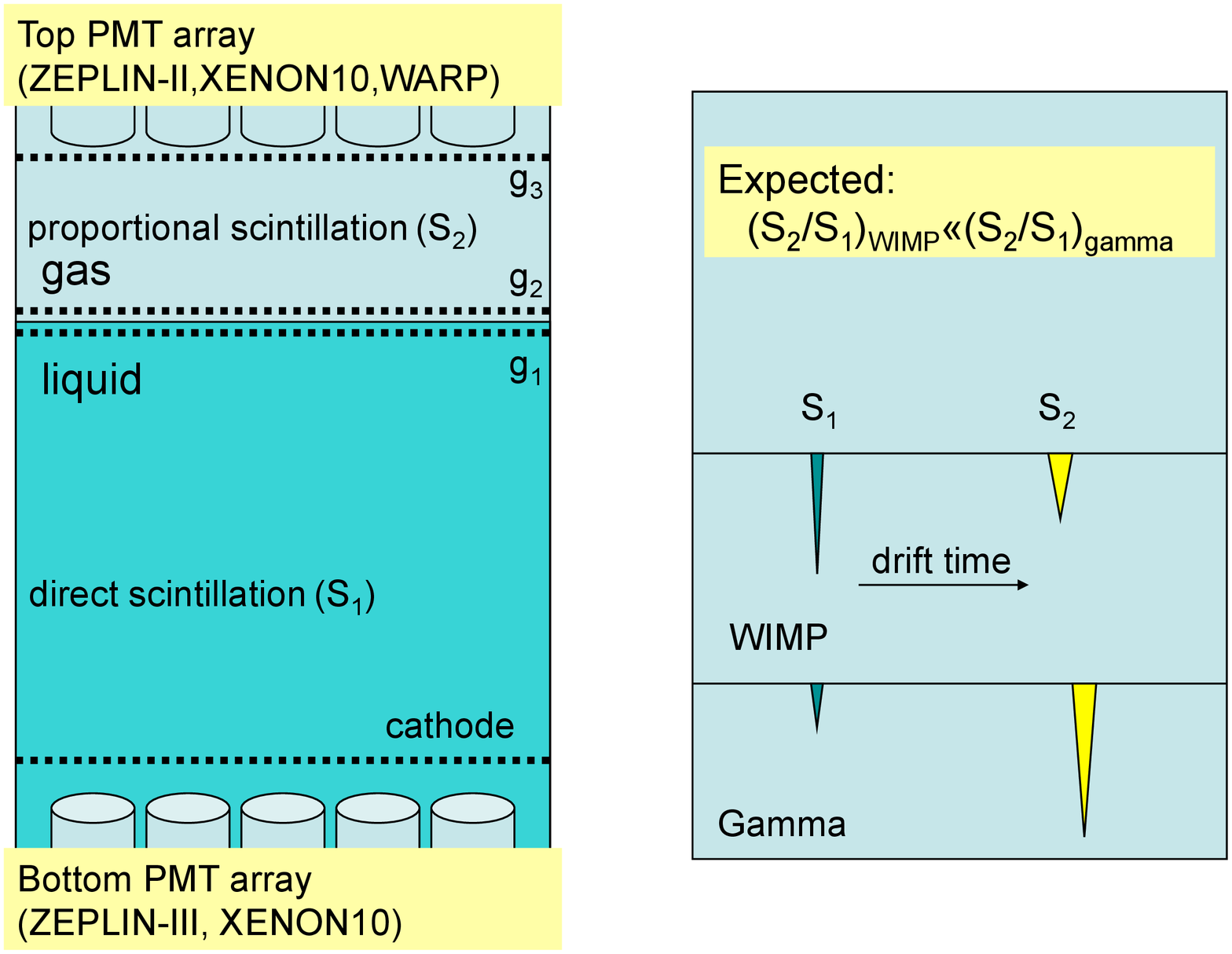}
\end{center}
\caption{Schematic drawing of a double phase detector (not to scale).
The grids g1, g2 and g3 allow the setting of the three electric fields present in the
double phase detectors as described in the text. The varying configurations of 
the photomultipliers, simplified in this diagram, are shown in Table \ref{tab1}. 
S1 is the prompt scintillation signal and S2 is the 
delayed proportional one (see text). Typical time differences between the
two signals are in the $\mu$s range (drift time, see Table \ref{tab1}). 
The measured signal ratio S2/S1 depends on the configuration of the apparatus, 
 the energy and the 
nature of the particle involved (c.f.r. Fig.  \ref{dis} and \ref{fff}). }
\label{schema}
\end{figure}

The  two phase detectors based on noble gases were originally proposed many years ago \cite{Dol}  
 and just
 recently reconsidered as part of different projects \cite{bolz,Cli2,Suz,Yam,Xpro}.  
The main aim and function of the dual phase (li\-quid/gas) operation is to measure  the direct 
scintillation in the liquid and ionisation, via proportional scintillation in the gas (see Fig. \ref{schema}). 
In fact electrons produced by the radiation interaction in the liquid  can be accelerated by an electric field in the gas to
acquire enough energy
to excite molecular states \cite{Suz2} of the gas and produce  
secondary scintillation.
Therefore the performance of the detector with regard to the scintillation response in the liquid and gas phase 
and in the presence of electric fields has to be known
as well as the efficiency of the processes. Due to the far UV light collection special 
photomultipliers (PMTs) and in some cases wave-length shifters have to be used.
It is important to know the quantum efficiencies of the PMTs used not only at UV light emission but also at cryogenic 
temperatures and to be able to overcome the difficulties linked to monitoring 
gain and stability of very low energy at these temperatures.

The main reason that this kind of detector is used in WIMP investigation is that
different particle species are expected to produce different amount of
 scintillation light both for the prompt scintillation signal (S1) in the liquid and the delayed proportional one (S2)  in the gas, with respect to
the same amount of nuclear recoil equivalent energy. Therefore
the ratio between the two signals is the main parameter used to discriminate between nuclear 
recoil events expected after WIMP-nucleus elastic scattering, 
and electromagnetic events. As we mentioned earlier, the electromagnetic component of the counting rate in these detectors is ``a priori'' considered
background and statistically rejected following the procedures described in Sect. \ref{Rej}.

Let us now turn our attention to some general aspects of the electric fields. 
Three electric fields are present in the two phase mode operation, acting in three different regions: the drift field is set between
 the cathode and a first grid (g1), placed just below the liquid level. A second grid (g2) is set just above the liquid level, in the gas.
These two grids are used to increase the field in the transition region between the two phases.
The extraction of electrons from the liquid is a delicate phase and a higher electric field is needed \cite{Dol} with respect to that applied to the liquid. 
 Extraction efficiency is different for each noble gas liquid, for instance the extraction efficiency for argon
 is higher than that for liquid xenon (for xenon a field  of 5 kV/cm is needed to  have about 90\% \cite{Gus}),
 which means that different extraction fields are necessary for comparable values of extraction efficiency.
 Complete stability of the electric field and the liquid surface are obviously needed and should be demonstrated at a 
 level to avoid systematic effects. 
The third grid (g3) is placed in the gas and the corresponding electroluminescence field accelerates electrons in the gas gap to produce
proportional scintillation. The proportional scintillation process in noble gases is is well known \cite{Suz2}. 
Studies have been performed on the light yield as a function of the electric field and gas pressure \cite{Fav} mainly for xenon.
For a given pressure the light yield is described by a linear function of the electric field E, i.e. roughly L = s (E - E$_0$), with s and  E$_0$ being functions of the pressure.
 The parameter s depends on the efficiency of light production and detection and can vary from
experiment to experiment. 
The process has a threshold which is a characteristic
value of the reduced electric field E$_0$/p (the electric field divided by the gas pressure) and has been reported to range from 0.84 kVcm$^{-1}$bar$^{-1}$ \cite{Fav} to
1.3 kVcm$^{-1}$bar$^{-1}$ \cite{Fei} for xenon, and from 0.5 to 0.7 kVcm$^{-1}$bar$^{-1}$ \cite{Mon1,Dia} for argon. 
An excitation efficiency, i.e. the fraction of energy that the electron acquires from the electric field and uses to produce excited states, should be reported for each apparatus. This efficiency is obviously
dependent on the E/p value applied and decreases rapidly below a critical value (estimated to be around 2 kVcm$^{-1}$bar$^{-1}$ for xenon \cite{Mon}).
 In addition, the wavelength of the emission spectra of proportional scintillation in noble gases also
depends on gas pressure \cite{Tak1}. 
A fundamental quantity that should be reported is the electroluminescence yield, i.e. the number of secondary scintillation photons
produced per drifting electron per unit path length. It is difficult to obtain an absolute measurement for this parameter, but it is particularly important as  
it characterizes the response of the detector. The values obtained and reported for xenon, for example, are not in agreement  \cite{Ngo,psci,Apr10,Fon,Mon} 
and are generally worse than those evaluated 
using Monte Carlo and Boltzmann calculations (only available for configurations at room tem\-pe\-ra\-ture) \cite{sim}.
The difference in these measurements is discussed in depth in ref.  \cite{Mon}. In fact one can define the reduced 
electroluminescence yield as the electroluminescence yield divided by the pressure, Y/p (photons electron$^{-1}$ cm$^{-1}$ bar$^{-1}$). 
The reduced electroluminescence yield shows an approximately linear variation with the reduced electric field E/p, and 
the scintillation amplification parameter (i.e. the slope of this linear dependence reported in Table 1 of ref. \cite{Mon})
 varies from 70 photons electron$^{-1}$ kV$^{-1}$ to 140 photons electron$^{-1}$ kV$^{-1}$.
The difference in these values are attributed to different gas purity \cite{Mon}. The most recent measurements \cite{Apr10,Fon,Mon} are more in agreement but only
two are performed at cryogenic temperatures \cite{Apr10,Fon} and one group \cite{Fon} found
a dependence of the obtained photoelectrons/keV on the amount of xenon  present in the liquid phase close to the saturated vapour pressure (considering that
 the system is defined by liquid xenon +  xenon vapour). This could be an important 
aspect for the working conditions of double phase experiments.\\
When the electric field increases and  the secondary ionisation occurs, the ener\-gy resolution of the detector deteriorates and there is a non-linearity in the electroluminescence
 yield due to electroluminescence produced by se\-con\-dary electrons. 
  The reduced field corresponding to the ionisation threshold for argon has been reported by ref. \cite{Mon1} as 2.8  kVcm$^{-1}$bar$^{-1}$ and by ref.  \cite{Dia} as 3.8 kVcm$^{-1}$bar$^{-1}$ while 
  the corresponding value for xenon has been reported to be  8 kVcm$^{-1}$bar$^{-1}$ by ref. \cite{Mon}, 5 kVcm$^{-1}$bar$^{-1}$ by ref. \cite{Fav} and even about 4 kVcm$^{-1}$bar$^{-1}$ by ref. \cite{Ngo}. 

The main characteristics of dual phase ZEPLIN-II, ZEPLIN-III, XENON10 and WARP detectors 
are summarised in Table \ref{tab1}. Let us now look in more detail at the apparata.

\begin{table}[htbp]
\caption{Main features of ZEPLIN-II, ZEPLIN-III,  XENON10 and WARP.}
\begin{center}
\resizebox{\textwidth}{!}{
\begin{tabular}{llllll}
\hline
   ~                      & ZEPLIN-II \cite{zep3}        &ZEPLIN-III \cite{zep5,zeplas}       & XENON10 \cite{xen1,fio,Apr9,Ober}   &             WARP \cite{arg2}  \\
  \hline
   site                   & Boulby, UK                          & Boulby, UK                     & LNGS                               &          LNGS                      \\
  detector           &   ~                                           & TPC                                      & TPC                                 &             drift chamber          \\
  gas                   &  xenon                                    & xenon                            & xenon                           &   argon                                  \\
sensitive mass & 31 kg                                     &   $\simeq$50 kg                              & 15 kg                               &               2.6 kg                     \\
fiducial mass    &    7.2   kg                                &   6.5 kg                               & 5.4 kg                              &               1.83 kg                 \\
n. PMTs             &      7 top                                &  31 bottom                     & 41 bottom                                &          7 top      \\
     ~                      &     ~                                      &      ~                                 & + 48 top                                   &     ~                                     \\
PMT size  &  13 cm $\oslash$                              &   5.08 cm $\oslash$   & 2.54 cm                           &     5.08 cm $\oslash$      \\
drift field             &   1 kV/cm                           &    3.9 kV/cm$^*$         & 0.73 kV/cm                    &     1 kV/cm                       \\
extraction field   &   4.2 kV/cm                        &           --               &     5 kV/cm & 4.4 kV/cm \\
field in gas         &  8.4 kV/cm                          &        7.8 kV/cm$^{@}$            &        9 kV/cm &      1 kV/cm           \\
reduced field$^{**}$ &  $\simeq$5.6  & $\simeq$5 & $\simeq$4   & $\simeq$1.1 &\\
 (E/p kVcm$^{-1}$bar$^{-1}$) &                               &                                         &                                                &    \\
depth of liquid   &     14 cm                                 &         3.5 cm                        & 15 cm                              &      7.5 cm                        \\
depth of gas      &                     --                            &        0.5 cm                        &          $\simeq$1.2 cm       &     8.75 cm                   \\
max. drift time    &      73 $\mu$s                          &      17$\mu$s                   & 75$\mu$s                         &     40$\mu$s                    \\
resolution in z  &     --                                            &  50 $\mu$m$^{***}$              & $<$ 1 mm                       &       --                                  \\
x-y resolution     &             $\simeq$1 cm          &  sub-cm$^{***}$                      & few mm                            &         --      \\
wave-length shifter        &             no                                     &    no                                 &   no                                     &    reflector + TPB      \\
reflector  &     PTFE                                                &                 no                        & PTFE                               &                   \\
\hline

$^*$ (1.3 --8.9) kV/cm and $^{@}$ 18 kV/cm by design \cite{zep5}\\
$^{**}$ deduced.\\
$^{***}$by design \cite{zep5}\\
\end{tabular}}
\end{center}
\label{tab1}
\end{table}

In the different experiments, the time difference between S1 and S2 signals, in general, determines the drift 
time of the charge, so it 
is used to provide the event depth, i.e. the z-coordinate, while the horizontal event definition (x-y coordinate) is
derived from the hit pattern of the PMTs.
The spatial
resolutions obtained are very different and depend on applied trigger conditions, the size and numbers of PMTs, and,
 when given, have been included in Table \ref{tab1}. 
This spatial information is used for some data reduction which is discussed in Sect. \ref{Rej} and \ref{Ana}. 
The values reported by different groups for the depth of the liquid illustrated in Table \ref{tab1} correspond to the height of the drift volume 
and not to the real depth of the liquid that can vary greatly. For example the configuration chosen by ZEPLIN-III 
is 25 cm, while the drift depth is 3.5 cm.

XENON10 and ZEPLIN-III are defined by their authors as dual phase time projection chambers (XeTPC)  because they were designed to identify
 3-D interaction positions.
  
XENON10 and  ZEPLIN-II are conceptually similar in design. 
The main difference between the two installations is the presence in XENON10 of
two arrays of photomultipliers (89 in total) one at the bottom immersed in the liquid and the other at the top in the gas. These PMTs detect 
both direct (S1) and proportional (S2)
scintillation light, while there is only one array of PMTs  in ZEPLIN-II set in the gas for both the measurements.

 The top array of PMTs in the XENON10 provides the hit pattern of the events.  
 The ZEPLIN-II authors do not report the resolution obtained for the z-coordinate.
  The authors discuss the horizontal event definition, i.e. the x-y event location, 
 evaluated using the relative pulse areas from S2 signals.
 Obviously uncertainty arises due to the large PMT size and the small
 photon statistics at low energy (see Sect. \ref{Cal}).

The WARP group mentions a position reconstruction of the event in the z-coordinate and the x-y-coordinate but no information is given on
the correspon\-ding resolution.  They add that future technology will allow 1 cm accuracy for the larger detector under construction \cite{arg2}
but even this future goal does not appear ambitious when compared to
the spatial resolution already achieved by the other set-ups (see Table \ref{tab1}).

Recently ZEPLIN--III is operative at the Boulby Underground Laboratory in the U.K..
 Surprisingly the size of the
detector has been reduced and a more segmented configuration in the x-y view favoured, making it similar to XENON10 but with very thin layers of liquid and
gas. This has probably been done to overcome some of the drawbacks that have emerged in all of the installations. 
These drawbacks will become clearer as we proceed. In this apparatus a single pair of outer electrodes is used to produce all three electric fields.
No information is given on the sensitive volume and only the fiducial mass is reported. A 50 kg stock of xenon is planned for 6.5 kg of fiducial mass, with
 an active volume of about 12 kg liquid xenon above PMTs.
However, their confidence in this new design, to improve light collection and energy threshold, appears to be difficult given what the same authors say
in their papers\cite{zep6,zeplas} (see Sect. \ref{Cal},\ref{Qf},\ref{Ana}).

As a general observation, the definition of an inner fiducial volume should be considered with the utmost care in the Dark Matter detection. 
 In fact, there are many factors (see Sect. \ref{Uni})
 that can affect the reliability of the fiducial volume in the field of Dark Matter due to the very low energy of the processes involved.

The 2.3 l WARP prototype actually has only a 1.87 l 
sensitive volume (while the fiducial one is 1.32 l corresponding to 1.83 kg) and works as a
two phase argon drift chamber. 
The detector has a top array of PMTs  placed in the gas phase and the walls are coated with  an organic 
wavelength shifter, Tetra-Phenil-Butadiene (TPB) \cite{arg1}. 
 In ref. \cite{arg2} it is specified that the PMT windows are also coated with TPB.
 The inner detector volume has the reflector glued on a Mylar TM sheet with TPB deposited
 on its active side.  Hence about 95\% of 
the surface surrounding the active volume is covered \cite{arg2}.
The PMT's quantum efficiency at the emission wavelength of TPB is reported by the authors to be about 18\% and the reflectivity in the TPB emission 
spectral region was measured to be about 95\% \cite{arg1,arg2}. No mention is made of the thickness, fluorescence decay  time,
conversion efficiency at the wavelengths of interest and mechanical stability of the TPB. 

None of the experiments considered here provided any information concerning the parameters resulting from
  the applied electric fields,  that are so important
for characterising the proportional scintillation process for each specific detector. Indeed the only further information that was added to that in Table \ref{tab1},
is summarised below. 

The WARP authors state that the PMT's windows are 4 cm over the g3 grid and that  the field allows electron collection on g3. 
The liquid-gas interface is placed 0.5 cm over g1.

For XENON10 it is stated that two of the four stainless steel mesh electrodes are set
in the liquid and the other two in the gas and the bias voltages are appropriate \cite{xen1}.
The bottom PMT array is 1.5 cm below the cathode mesh.

ZEPLIN-II added they had obtained a
90\% extraction efficiency of 
electrons from the liquid surface and a secondary yield of 
$\sim$230 electroluminescence photons per extracted electron from the liquid surface, at 
the mean operating pressure of 1.5 bar \cite{zep3}. No information is given
on the gas gap,  one of the most important parameters for determining the proportional scintillation signal.

A consistently higher value of the drift field is reported in ZEPLIN-III.

The reported values for the reduced field corresponding to
 the production of secondary electrons by xenon could create problems for ZEPLIN-III and ZEPLIN-II, 
 as the reduced field in the gas is in this range or higher. In ge\-ne\-ral,  
  this value should be measured and reported  for each apparatus as it is strictly dependent on experimental conditions.
 We will not go any further into this argument as no information is available concerning the experimental set-ups for the entire running periods.

Taking again xenon as an example, the question concerning the
 dependence of reduced electroluminescence yield (the number of secondary scintillation photons
 produced per drifting electron per unit path length and unit pressure)  for proportional scintillation on gas pressure is still open to debate.
 Some authors have found this dependence \cite{Pav}, others have not and attributed this behaviour to the different UV
 spectral response of the detectors used \cite{Fav}, i.e. light production and detection efficiency. This information is absent in 
 papers on the dual phase detectors available so far, while it merits an accurate study and should be reported 
 for all the detectors used to publish results on Dark Matter.

Finally the techniques regarding the application of liquid neon are, in principle, even more complex than those involving other condensed rare gases,
due to the cryogenics (the neon boiling point is at 27 K) as well as 
the many aspects of far UV light detection. Recently liquid neon has been proposed for WIMP and solar neutrino
flux investigation, but in a single phase detector without electric fields, and by studying the application of pulse shape discrimination techniques  \cite{Mac1}.   
At present however the use of liquid neon in this field is still at a very early stage, despite the multi-tons volumes already claimed for the
future. 

\section{Purity and radiopurity}

The WIMP direct detection approach pursued by these experimental techniques
relies on low background apparatus. This is particularly true when considering that no model independent signature can be exploited for the
 detection\footnote{The exploitation of a model independent signature in itself  acts as an effective
background rejection, with respect to the low energy application of many data selections and  discrimination procedures  that always have a statistical nature, decide 
``a priori'' the nature of the interaction and leave full uncertainty on the final nature of the selected events, for instance nuclear recoils with respect to end-range 
$\alpha$'s etc.}.
 As has been documented by long research activity in 
rare event physics, low-background techniques require very long and accurate work, and efforts are focussed on reducing standard contaminants such
  as $^{238}$U and $^{232}$Th chains and $^{40}$K due to notable presence in nature. Non-standard contaminants
  should also be investigated in depth (see e.g. ref. \cite{Heu}). Many years of dedicated efforts and  considerable experience are needed to obtain several
  orders of magnitude of counting rate reduction, as was the case for example for
  the Ge ionising experiments that took several decades (see e.g.  ref. \cite{Avi}).
  Therefore let us examine two phase
  noble detectors from this point of view.

In the case of liquid noble gases
scintillation light is emitted in the far-UV region.
Therefore efficient light transmission and collection is fundamental. Impurities (such as nitrogen, water, hydrocarbons, etc.) 
can absorb UV light and act as inefficient wave-length shifter or as quenchers degrading light collection \cite{Mey}. 
When scintillation light is absorbed by some impurities the consequent overlap of the absorption and emission light could be the cause of a
position dependence of the detector response already observed in the past \cite{Mih,Bald}.
This phenomenon also affects, to varying degrees, the performances
of dual phase detectors discussed in this paper (see Sect. \ref{Uni}). This kind of dependence is particularly dangerous  when considering large
acceptance detectors. 
The choice of the materials is also critical because the most of them are UV absorbing or have poor reflectivity for UV light.

Let us consider liquid xenon. In general light attenuation consists of two components \cite{Bald}:
a scattering length and an absorption length. Little information is available on scattering length \cite{Sei,Nev} while it is known that absorption length is 
strongly dependent on the reflective and purity properties of the materials and  gas used. The total attenuation length is the measured
quantity and published values range from cm to m \cite{Bar,Apral,Xel,Ishi,Nev,Bald}.

In addition due to the presence of electric fields (and therefore taking into consideration the importance of 
efficient propagation, extraction and collection of drifting electrons) impurities that affect ionisation collection must be avoided.
 The most dangerous are the electro-affine ones (SF$_6$, O$_2$, N$_2$O, CO, CO$_2$, H$_2$O) \cite{Bak}.
In the range of work of double phase detectors (0.73 kV/cm in XENON10 and 1 kV/cm in ZEPLIN-II and WARP) 
the presence of SF$_6$ is critical because the rate constant for electron attachment is very high  (about 10$^3$ higher than O$_2$) while the rate constants
in liquid xenon for electron attachment to SF$_6$, N$_2$O and O$_2$ are slightly higher than in liquid argon. In addition, the rate constant for electron attachment to N$_2$O
greatly increases
with increasing field strength: one order of magnitude in the interval from 1 kV/cm to 10 kV/cm (the working range of ZEPLIN-III) \cite{Bak}.

It has taken over two decades to reduce electronegative and scintillation absorbing impurities to levels
 around $\simeq$ 1 ppb \cite{Lope} and even this is not yet enough to achieve the best detector performances 
 predicted by theoretical model evaluations \cite{Lope}. This is why R\&D efforts are continually striving to achieve higher purities.

Therefore with liquid noble gases, high level gas and material purity is required, as well as 
a specific purification system whilst high temperature
degassing of the detector and purification/filling/recovery line under an ultra-high vacuum is essential. Achieving a sufficient level of purity is however only 
the first step, the second is maintaining it over the data taking period and being able to guarantee the
 reproducibility of the measurement i.e. being able to achieve the same purity levels for each new liquefaction.
  The importance of these steps should not be underestimated,  even when the noble gas is recovered, because
the risk is that the detector is ``rebuilt each time'' with different experimental parameters and performances although the apparatus and the procedures followed
are the same.

Liquid xenon has a larger electron diffusion coefficient than liquid argon \cite{Dok1} and a higher boiling point
and this creates further restrictions for purification and the materials used.

The use of a detector and very low radioactivity materials in the search for rare events in Dark Matter investigations requires ad hoc solutions
that take into account purity, radiopurity and the optimisation of working conditions of the detector in general and in light of its response at very low energy.

Contamination, for example from $^{222}$Rn (T$_{1/2}$= 3.82 days) and $^{220}$Rn (T$_{1/2}$ = 56 s) isotopes
and their daughters,  from 
$^{238}$U and $^{232}$Th chains has to be avoided. This is done by selecting appropriate materials and putting protocols 
in place that protect the detector surface,  the filling/purification/recovery line,  the shields etc. from pollution\footnote{It is essential to use
 very low radioactivity materials and avoid pollution such as e.g. from
 $^{222}$Rn from the external environment for example as $^{222}$Rn would increase the measured counting
  rate and could also mimic, by end-range alphas, candidate recoils (as can neutrons, the fission induced events, etc.).}. Some quantitative estimation by authors 
  of double phase detectors applied to Dark Matter detection could be useful.

This preventive action is important not only during the running time but also in the preceding phases of transportation, storage and assembly of the apparatus.
So walking inside the shields without the proper equipment,
as can be seen in some photos, or leaving  the electronic systems and detector installation in an open-air  environment should be avoided.

\subsection{Gas}

The  double phase experiments considered in this paper use the natural isotopic composition of Noble gases.
 Due to the very short emission wavelength of argon and neon a wave-length shifter is used. 
We begin by considering the purity problem from the isotopic composition of the different gases.

All naturally occurring isotopes of xenon are non-radioactive; xenon radioactive isotopes only have very short 
half-lives. The longest half-life isotope is $^{127}$Xe (36.4 day). 
In any event, their possible cosmogenic presence can be removed by a suitable period of storage underground.  
 XENON team discussing possible Xe radioactive contaminants has stated \cite{xeba}:
 {\it Natural xenon does not contain any long-lived radioactive isotopes 
(apart from the double beta emitter $^{136}$Xe which has a 
half-life larger than 0.5$\cdot$10$^{21}$years)} or in the proposal \cite{xepr}: {\it According to our simulations, the 
contributions to the electron recoil background from the cosmogenic activation of Xe and the 
double beta decay of $^{136}$Xe are negligible compared to other components}.
Considering that  the process of double beta decay in $^{136}$Xe has not yet been observed and the existing limits on the half-life are in the 10$^{22}$ y 
range this contribution is obviously negligible with respect to other well known contaminants.

Natural neon has no radioactive isotopes. Its radioisotopes 
have very short half-life and  the same considerations made above for xenon also hold true for neon, including concerns about the  radioactivity of $^{136}$Xe:
{\it Neon has no inherent radioactivity, unlike argon, krypton and xenon which have natural radioactivity 
from the isotopes $^{39}$Ar, $^{85}$Kr and $^{136}$Xe} \cite{Har,Mck}.

Natural argon also has no radioactive isotopes. Its longest long-lived radioisotopes are:
$^{37}$Ar (half-life= 35 day, EC),  $^{42}$Ar (half-life=33 y, $\beta^-$ emitter) and $^{39}$Ar (half-life= 269 y,  $\beta^-$ emitter). 
The $^{39}$Ar and $^{42}$Ar, can be produced by cosmic ray interaction, neutron interaction with 
stable argon isotopes and spallation reactions and they have long been considered potential sources of background in rare event searches
(see for instance \cite{BarCen}).  $^{42}$Ar was studied by different authors and the most restrictive limit 
was set \cite{Ash} at less than 6 $\cdot$ 10$^{-21}$ parts of $^{42}$Ar per part of $^{nat}$Ar at the 90\% C.L. that corresponds
to less than 85 $\mu$Bq/liter in liquid argon \cite{arg1}. The radioactivity of $^{39}$Ar, 
on the other hand, is more significant. Its activity in the atmosphere is known to be about 10 mBq/m$^3$ \cite{Loo}.
The specific activity in natural argon was also measured recently by WARP \cite{arg1}:  
 1 Bq per kg of $^{nat}$Ar and the value is consistent with the previous determination. 
 
 The majority of authors do not describe the starting quality of the used gases, the residual contaminants after the purification, the quality of the
  filling/purification/recovery line, the Ultra-High vacuum system etc. in detail and a lot of experimental information is missing; in some cases the reader is referred to future papers.
However this information is important because, for example, the realistically reachable purity is a function of its initial level and of the nature of the original impurities.
 Furthermore the level of some specific impurities plays a significant role in the reachable light and charge performance of liquid noble gas detectors (see literature)
 and lastly, impurities can also be released in the gas/liquid during ''purification'' phase, due to contamination in the purifier materials
  themselves (e.g. OXISORB powders \cite{noi1}) and/or during operation in which the surfaces of materials come into contact with the gas/liquid.  

The liquid argon in the chamber used in WARP is argon 6.0 
supplied by Rivoira S.p.A.. It is purified from electronegative impurities and continuously recirculated through 
a chemical filter:
{\it equivalent contamination of less than 0.1 ppb of O$_2$ by 
using the chemical filter Hopkalit$^{TM}$ from Air Liquide} \cite{arg1}. However they also state ref. \cite{arg2}:
 {\it The detector has been filled with good grade commercial 
argon, without any ``ad hoc'' additional purification} and even if an argon recirculating
system used for purification is mentioned, the residual concentration of electronegative
impurities in argon is reported as being {\it of the order of 
$\leq$1 ppb (O$_2$ equiv.)}. See also ref. \cite{wao} for recent tests.

The Xe gas used for XENON10 was commercially procured and from the 
electron lifetime of  2.0 $\pm$0.4 ms  inferred from calibration data, they derive a purity limit of
$<<$1 ppb (O$_2$ equiv.) \cite{xen1}.
They also state  that a single high temperature
getter and a closed circulation system \cite{Apr9} are used to purify and keep the liquid xenon clean but without giving any further details \cite{xen1}.  

In the case of ZEPLIN-II experiment, xenon is recirculated 
{\it by drawing liquid from outside the active volume using an internal heater and 
a Tokyo Garasu Kikai MX-808-ST diaphragm pump [25], distil\-ling through 
a SAES getter PS11-MC500 purifier} \cite{zep3}. A flow rate of 3 SLPM (Standard Liters Per Minute)
is maintained to ensure a purity of xenon corresponding to $>$100 $\mu$s \cite{zep3}. 

To give an example of problems that could arise as a result of this situation let us consider the value of the light absorption cross section of water.
In the UV range of xenon and argon scintillation light,
the light absorption cross section of water has the largest value (see Fig. 4(a) in ref. \cite{Bald}). 
However it would appear unlikely that all the PMTs have been heated to high temperatures
 (generally it is done at more than 150 $^{0}$C)
for a long time in ultra high vacuum, to remove water impurities present not only in the original gas
but also desorbed by all the materials in contact with it.
So, strictly speaking, the gas purification mentioned by authors only refers to  Oxygen levels.
 Furthermore no mention is made about possible water traps in the line, the cleaning of the plant and detector, degassing of materials 
like meshes and wires, fittings, o-rings  etc. that are in direct contact with the liquid and gas for long periods of time.

The radioisotope $^{85}$Kr is present in liquid noble gases. It is a beta emitter with an endpoint
 of 687 keV and half-life of 10.76 y.
The presence of this radioisotope in the atmosphere is mainly of anthropogenic nature, in fact natural
 activity of $^{85}$Kr
is negligible from spontaneous fission of Uranium and Thorium and from neutron activation, $^{84}$Kr (n,$\gamma$) $^{85}$Kr. 
However since the 1950s increasing levels of $^{85}$Kr have been recorded (25 mBq m$^{-3}$ y$^{-1}$ \cite{Sid})
as a result of Plutonium reprocessing for nuclear reactors.
Its concentration ratio is $^{85}$Kr/Kr
$\simeq$ 10$^{-11}$ \cite{Los}.

The ultracentrifugation procedure used to ``enrich'' or ``clean''  a gas like xenon is expensive but very effective in
eliminating most of the $^{85}$Kr present and can be considered an important tool for cleaning and removing 
 low le\-ve\-ls of gas impurities. The ``cleaning'' procedure does not modify the isotopic composition.
For example, xenon enrichment in $^{136}$Xe \cite{Bara,Vui,Bel} and in $^{129}$Xe \cite{noi1}: the enrichment in
 $^{129}$Xe at 99.5\% certified by ISOTEC and checked upon arrival by mass spectrography \cite{noi} corresponds to levels 
of Kr $<$ 20ppb, while for the gas enriched in $^{136}$Xe at 68.8\% by Oak Ridge a limit of 3.2 $\cdot$10$^{-19}$
atoms ($^{85}$Kr)/atom(Xe) at 95\% C.L. was recorded \cite{Bel1}.
For comparison purposes it is worth recalling that natural, commercially available Xe of 99.998\% purity typically
 contains $>$ 400 times higher Kr concentrations.

As WARP does not recover the argon used the level of contamination may vary considerably in subsequent runs
and this is true for all the liquid noble gases.
The measurement performed by the 2.3 l prototype  
gives (0.16$\pm$0.13) Bq/liter for  the $^{85}$Kr  contamination. 
In any case the uncertainties that have arisen concerning  
the determination of  $^{85}$Kr and  $^{39}$Ar content are due to the definition of the sensitive volume and are
 based on the
different energy end points. The latter definition relies on energy calibration and energy resolution; the authors quote
systematic errors \cite{arg1}.

ZEPLIN-II states that the Kr contamination is below
the 30-40 ppb level, quoting the manufacturerÕs specifications \cite{zep3} and no purification on Kr is performed.

The manufacturer used by XENON10 uses 
a Kr cleaning procedure,  and XENON10 declares that the gas used has 
{\it  a guaranteed Kr level below 10 part per billion (ppb), from repeated passages
 through a dedicated cryogenic distillation column} \cite{xen1}.

50 kg of xenon collected between 20--40 years ago from ITEP, from an underground site source, is used in ZEPLIN-III \cite{zep5}. The Kr content
is reported to be $\simeq$5 ppb, but the contribution of low energy background due to $^{85}$Kr is about 1\% of the background produced from
the PMTs presently used.

Improvements are planned for xenon for all the future projects but have yet to be tested: 
levels of few 100 ppt of $^{85}$Kr for xenon are declared achievable in a 100 kg Dark Matter detector by using distillation (XMASS) \cite{xma}
or adsorption on a charcoal column \cite{apr}; for argon further purification is planned as well as an active anti-coincidence external shield made from liquid Ar
(but what about the residual contamination from the internal reflector+TPB?).
Two purification methods are under study for argon: one using centrifugation, possibly giving a residual contamination for $^{39}$Ar of less than 2\%
with respect to natural content \cite{arg2} and the other method involving argon extraction from
an underground stacked gas reservoir.  This gas could be free from some impurities \cite{arg2}, but the presence of  $^{39}$Ar could
be significant even in geological argon.  In fact, the effect of cosmic rays  on geological argon is related to the depth of the argon source and
$^{39}$Ar can also be produced by a number of reactions in the {\it subsurface} \cite{Gal08},
 such as $^{39}$K(n,p)$^{39}$Ar; its presence also depends on local U, Th and K concentrations \cite{Gal08}.

Results from tests that have already been performed on batches of  depleted $^{39}$Ar argon in the 2.3 l WARP detector have not yet been published.
 Measurements are available in ref.  \cite{Gal08} of a depletion factor $\ge$ 10 at 97.7\%C.L.; 
 this implies $\le$ 0.1 Bq of $^{39}$Ar per kg of $^{nat}$Ar.

Possible contaminants of neon are $^{85}$Kr, $^{39}$Ar and tritium, studies on their removal have been considered \cite{Mac3} but no quantitative
analysis has yet been published.

\subsection{Installation}

We now look at the purity and radiopurity of the different detectors and in particular the installations, materials and shields.
Once again little information is given considering the number of materials inside the detectors: PMTs,
feedthroughs, mesh and wire structures, reflectors or wavelength shifters etc.

Freshly produced, pre-Cernobyl electrolytic copper is generally preferred for its good radiopurity and
 in cryogenics the thermal contacts are usually made 
 of Indium; this cannot be used in the field of Dark Matter owing
 to the presence of $^{115}$In (a beta emitter with endpoint at 482 keV). So, it is difficult to manage 
 copper (of the vessel) and feedthroughs (stainless steel) seals.
 These difficulties generally lead to the choice of 
 stainless steel at a preliminary stage,
 as in XENON10  despite the lower radiopurity
 whilst ZEPLIN-III opted for Indium.
 
The XENON10 was not specifically designed as a low background detector,
but the authors of the ZEPLIN-II state that it is primarily
constructed from low background materials even though a background population of Radon
progeny events was observed in the measurements they published \cite{zep3}.

Only the materials used in the PMTs were classified as selected
for their radiopurity in WARP. They quote the supplier's specifications 0.2 Bq/PMT, as total $\gamma$ activity above
100 keV, dominated by the $^{232}$Th and $^{238}$U chains \cite{arg1}.
In the paper \cite{arg2} the authors consider the presence of impurities useful
{\it in this preliminary phase of development of our novel 
technology} in order to develop the $\gamma$ and $\beta$ radiation rejection capability in short data taking time
and also for the contamination of Rn soon after filling: {\it The presence of such a Radon induced signal is 
very useful, since it has permitted a real time calibration 
of the scintillation light yield of heavy ion recoils} \cite{arg2}. In reality apart from an analysis of the energy spectrum of this
background above 100 keV in ref. \cite{arg1}, few plots on the corresponding S2/S1 
discrimination 
are given in ref. \cite{arg2}, which are needed to understand the nature of the different events.

Contaminations of  XENON10 PMTs have been measured \cite{APS}:
0.17$\pm$0.04 mBq/PMT for U,  0.20$\pm$0.09 mBq/PMT for Th, 10$\pm$1 mBq/PMT for K and 0.56$\pm$0.05 mBq/PMT for Co.

In the design and commissioning paper for the ZEPLIN-III \cite{zep5} the importance of checking the inherent
radioactivity levels of the detector and the procedures used is continuously stressed.
The ZEPLIN-III collaboration performed a screening of PMT materials before manufacturing.
The measured levels of impurities were 250 ppb in U, 290 ppb in Th and 1350 ppm in K.  So the expected background
 using Monte Carlo simulation is 10 events/kg/day/keVee\footnote{keVee means keV electron equivalent.}  in the low energy region \cite{zep5} which is still high. 
  The replacement of these PMTs with others under development with a 30-fold reduction in their radioactivity is planned. 
 The vessel is OFHC copper (type C103), the welds have been minimised and when unavoidable the electron-beam welding method was used.  
 Stainless Steel is present (in the used flanges and feedthroughs) and 40 g of Indium ($\simeq$10 Bq) was placed inside the seals, distributed 
 along the electrical feedthroughs of the PMT dynodes and the bottom flange of the vessel.  The authors ensure the all round passive
 shield (30 cm thick polypropylene and 20 cm thick lead\cite{zeplas}) is sufficient to prevent background in the fiducial volume (see later).

The WARP detector is shielded by 10 cm lead walls encapsulated in stainless steel boxes, plus polyethylene (an equivalent thickness of 60 cm) \cite{arg2}.
 No mention of polyethylene is made in the paper \cite{arg1}; in any case 
both papers \cite{arg2,arg1} report the same total counting rate.
The value reported for the measurement of ref. \cite{arg1} is 6 Hz (from the acquisition threshold of 40 keV up to 3 MeV) and
 the energy spectrum is shown from 100 keV (analysis threshold) up to 3 MeV. 
 In ref. \cite{arg2} no further specifications are given and the energy is always given as nuclear
recoil energy. In this last case the reported energy threshold used for the analysis is 40 keV of nuclear recoil energy without any mention of the
quenching factor. In conclusion, as the reader will surely have realised, this type of information is important, and must be complete in order to assess the detector under working conditions.

Two main background components were found by the authors in the WARP prototype \cite{arg1}: 1) contaminants present in
the materials surrounding liquid argon, dewar etc.
from $^{238}$U chain ($^{222}$Rn was present in the liquid argon used for the bath), $^{232}$Th chain, $^{60}$Co and $^{40}$K;
2) contaminants in the liquid argon target - mainly
$^{222}$Rn, $^{39}$Ar and $^{85}$Kr.
After filling the chamber with new argon the total decay rate of $^{222}$Rn is  around 1-2 Hz and is quoted to be a few tens of events/day
four weeks after the filling (considering the $^{222}$Rn half-life). Therefore production runs are started 
several weeks after each filling.  $^{14}$C is found in liquid argon, mostly close to the chamber walls  and has an estimated contribution of less than 50 mHz.
$^{60}$Co and $^{40}$K contamination is found in the surrounding materials
(the total energy 
spectrum measured and simulated is shown above 100 keV, but these 
last two contaminations are not  quantitatively given).

The reported background rate in XENON10 is 0.6 events/keVee/kg/day 
in the central
part of the detector and 3 events/keVee/kg/day near the edge 
\cite{xen1}, but although not stated in 
the paper, this rate is probably the value  obtained
after preliminary cuts, in fact  the raw rate shown in Fig. 6 of ref. \cite{Apb} reaches 10$^{3}$ events/keVee/kg/day.
The vessel and the cryostat are made from stainless steel, totalling 180 kg, around the liquid xenon, the 
feedthroughs are ceramic. A further 10 kg
of liquid xenon fills the area outside the sensitive volume. The shield is made from 20 cm of polyethylene and  
 20 cm of lead and nitrogen gas is flushed in the shield \cite{Apb}.

 The ZEPLIN-II shield is made up of 25 cm Pb and 
 an inner 30 cm hydrocarbon neutron shield, i.e. a liquid scintillator vessel
and a roof of solid hydrocarbon blocks, both with 
{\it Gd-layering} \cite{zep3}. The reported background rate - already corrected for the efficiencies - is 0.5 events/keVee/kg/day
 in the fiducial volume for the 5-20 keVee energy range. No information is given on the raw counting rates before selections.

\section{Thermodynamic working conditions}

In general stability and control of the thermodynamic 
working conditions are important for stable light collection, but
in a dual-phase detector, the thermodynamic stability is fundamental, given 
that the proportional light yield depends not only on the electric field in the gas, but also on the gas gap and the gas pressure \cite{Lan,Bol}.
The surface between 
the liquid and the gas phase must also be stable. 
We quote ref. \cite{sauli} pag. 491, on the liquid/gas surface: {\it Any heat source
 in the xenon such as ... can cause bubbling and thus reduced transparency (an unacceptable apparent partial energy loss)}, 
 and concerning the temperature gradients: {\it Although not completely understood, it has been experimentally observed
  that temperature stratification in the liquid xenon leads to apparent refractive index changes which
  could also cause light-collection losses.} 
  This book is on high energy Physics, obviously the case of very low energy applications in the Dark Matter field is far more critical.

Let us now consider how these aspects are taken into account
by the different experiments.

In XENON10, cooling is provided by a pulse tube refrigerator (PTR). The cryocooler, the signal and high-voltage feedthroughs 
are placed inside the shield with all the known drawbacks.
 These parts will be placed externally in the XENON100 apparatus. 
Only XENON10 has presented data on temperature stability for the
detector system, working at T =180 K and at about 2.3 atm, declaring a temperature stability  
$|\Delta T| < $ 0.005 $^0$C,
and a pressure sta\-bi\-lity $|\Delta p| < $ 0.006 atm for over ten months of operation \cite{Apr9}.
Under these conditions the fluctuations of the PMT gains are quoted $<$ 2\%. Unfortunately some information is missing, 
for example the number of temperature sensors used, the uniformity of temperature along the detector in light of the fact that
the vessel is in stainless steel, a material that does not have good thermal conductivity\footnote{The thermal conductivity  (see e.g. ref. \cite{term1})
is: i) 400 Wm$^{-1}$K$^{-1}$ at 300 K and increases up to 482 Wm$^{-1}$K$^{-1}$ at 100 K for copper (a material usually 
chosen for vessels in cryogenics); ii) 18 Wm$^{-1}$K$^{-1}$ at 300 K and decreases up to 16 Wm$^{-1}$K$^{-1}$ at 100 K
 for stainless steel; iii) 0.108 Wm$^{-1}$K$^{-1}$ for liquid xenon \cite{term2}.}
 and that 89 PMTs are operative in the
liquid + gas volume. This question can also stand for the new XENON100 installation.

The cryogenic WARP system is described: it is stated that the detector is
 contained in a stainless steel vessel,
25 cm in diameter and 60 cm in height and it  is cooled down to 
 about 86.5 K by an external liquid argon bath.  The authors say that the installation ensures  
 an absolute pressure of a few mbar above the 
external atmospheric pressure \cite{arg2} in the inner 
container and
{\it Two liquid argon 
level meters allow the liquid level to be positioned in between 
the two lowest grids with a precision of about 0.5 mm} \cite{arg2}. 
Therefore the drift chamber is operated at Ar boiling point (86.7 K) and 950 mbar, 
1.399 g/cm$^3$(at the atmospheric pressure of LNGS):
{\it  The pressure of the gas phase on the top of the 
chamber is naturally equalized to the surrounding atmospheric pressure} \cite{arg1}.

In ZEPLIN-II a refrigerator is used (IGC PFC330 Polycold refrigerator) linked to
a copper liquefaction head within the target vessel:
{\it the target chamber and internal structures are cooled convectively and by xenon 'rain' 
from this liquefaction head} \cite{zep3}.

In addition to the features mentioned earlier, temperature stability is also important because it has been found that the drift velocity of electrons
 depends  on the temperature in liquid xenon \cite{Dok1,Bene}. This behaviour is inversely
proportional to temperature. 
Therefore such a dependence cannot be neglected if the aim of the measurement is a precise position measurement.
This holds for whatever experiment owing to the possible non-uniformity
 of temperature along the detector due to thermal gradients. For example, even in the case of XENON10, 
 that quotes a high temperature stability, information such as how many temperature sensors are used,
 their position etc. cannot be neglected as temperature stratification may exist.

In this kind of physical application, control of working conditions is critical, as the case of ZEPLIN-II highlights: 
a specific problem on the cooling power meant 
they had to correct the secondary electroluminescence signal. In fact they write  that 
although temperature is controlled on the liquefaction head, the variability in cooling power 
varied the target gas pressure and environment temperature during runs. 
Therefore they say:
{\it Accordingly, this varied the secondary electro-luminescence
photon production, directly by changing the electro-luminescence gas pressure
and indirectly by changing the xenon liquid level between the extraction grids,
and hence the electro-luminescence field and path length in the gas} \cite{zep3}.
Again this underlines how important the stability of the thermodynamical conditions is.
The working temperature is not explicitly mentioned, but the pressure is reported to be $\simeq$1.5 bar 
(so the temperature should be around 172-173 K).

The cooling of ZEPLIN-III is performed with about 200 l of liquid nitrogen \cite{zep5}.
In general when used for long-term operation, despite its simplicity, the liquid nitrogen cooling method 
 is very inefficient and wastes a lot of liquid nitrogen.
In addition, the systems using liquid nitrogen are more unstable and troublesome with  respect 
to the ones using a cryogenerator. In fact due to the difference in boiling temperatures of liquid xenon and liquid nitrogen,
it is difficult to assure temperature stability of liquid xenon.
However, the authors declare to have temperature control in the
critical points of the detector,  a temperature stability better than 0.5 $^0$C 
and a pressure stability to 2\% in normal operations \cite{zeplas}.
The control on the liquid level it is given with sub-mm accuracy in three locations \cite{zep6}, while the
working temperature is $\simeq$ 173 K.

\section{Trigger}

All the apparata have arrays of PMTs. As known, the equalisation of their response and the stability of
their gain are features that should be tested. 
In addition to the equalisation of the PMT gains the number of photoelectrons (phe) depend on:
the average energy required to generate a scintillation photon (W$_s$), the electric field, the reduction in yield due to the
 quantum efficiencies of the PMTs, the photon collection efficiency (depending on the detector design, attenuation length, optical
 transparency of meshes etc.).
The quantum efficiency of the PMT depends not only on wavelength, but also on temperature
and should be under the working conditions. Due to the selection procedures 
on the collected data used by the experiments considered here, it can be useful to compare trigger conditions and the information digitised about 
pulse shape and time. 
 Let us consider the different authors' comments.

For each PMT in WARP the integrated signal from the anode is passively split in two signals, with
relative amplitudes 1:10, that are sent to two 10 bit flash 
ADCs with 20 MHz sampling frequency. For each trigger the event structure and the pulse shapes
of the PMTs are recorded.
For the trigger, signals are derived 
from the 12th dynode of each PMT and then they are amplified 
and discriminated with a threshold of 1.5 photoelectrons. The conditions for  trigger are defined as 
$\geq$ 3 PMTs of at least 1.5 photoelectrons each, i.e. about 3.5 keV ion recoil energy.
The quoted trigger efficiency is close to unity 
above about 20 photoelectrons (16 keV ion recoil energy) \cite{arg2}.

ZEPLIN-II is the only experiment that gives detailed information on data acquisition and data reduction \cite{zep4}.
The signal of each PMT is passively split in two lines: one is used for the trigger and the other is
digitised for the waveform. This digitisation is performed with  ACQIRIS  DC265 digitisers (500 MSamples/s, 
150 MHz bandwidth).
The  summed output from the liquid scintillator veto is also digitised.
The second line is amplified and discriminated and the trigger condition is given by
five-fold coincidences between different PMTs at single photoelectron level.   
The authors explain that, due to the chamber design, 
the central PMT sees a larger signal 
(on average) than the outer PMTs and when
large amplitude signals occur
optical feedback in the target can be produced. This corresponds to many 
long noise pulses that can occur for tens of $\mu$s after the main S2 signal, therefore
they are vetoed to minimise the DAQ dead-time.
The effect is to significantly reduce the trigger rate by 60\% \cite{zep3,zep4}.
The authors consider this procedure useful because it reduces:
{\it data processing and data storage requirements} \cite{zep4}.
The quoted hardware efficiencies are shown later (see Fig. \ref{xcut3}).
Both S1 and S2 signal can trigger the data 
acquisition, so
the digitisers acquire data
$\pm$100 $\mu$s around the trigger pulse. To reduce random coincidences,
 in an off-line data analysis, three-fold coincidences among different
PMTs at a single photoelectron level were used to identify and parameterise the primary signal S1 \cite{zep4}.
Multiple scattering events and events for which S2 saturates were rejected
 by data analysis too.
 
 The high refractive index of liquid xenon with respect to the gas is also an important aspect. It was measured to be
1.7 at 170 nm, decreasing down 
to 1.4 at 360 nm (where liquid Xe is considered at 
173.2 K and 1 atm) \cite{Mass}. While  at 180 nm it measured 1.5655$\pm$0.0024$\pm$0.0078
for the xenon triple point \cite{Bark}. Due to this high value, much light from the liquid is certainly 
lost in the ZEPLIN-II due to the total reflection passing through the liquid gas interface, 
as it only has PMTs on the top of the
detector.

 All the signals of the 89 PMTs are digitised at 105 MHz in the XENON10 and
 the trigger is provided by  the S2 sum of 34 top array centre PMTs \cite{xen1}.
 The S2 hardware trigger threshold is 100 phe, corresponding to about 4 electrons
 extracted from the liquid
{\it which is the expected charge
from an event with less than 1 keV nuclear recoil equivalent energy} \cite{xen1}. 
The quantum efficiency is reported to be $>$ 20\% at 178 nm \cite{APS}.
No schematic drawing of the electronic chain is given or detailed information
concerning how PMTs are managed in electronics and trigger.

The PMTs in ZEPLIN-III are arranged in a closed
packed array in copper where they fit, but instead of bringing all the individual connectors
out there is a common high voltage supply and dynode distribution system.
The system of 16 copper plates used to make the internal PMT dynode 
interconnections is described in detail, but no information is given on the performances concerning noise,
crosstalk etc. when they work together. A qualification of PMT's features at low temperature has been carried out 
separately for each PMT \cite{zep7}.
The ACQIRIS  DC265 digitisers are also used in ZEPLIN-III but the PMT signals are digitised over a 
time interval of $\pm$ 18 $\mu$s around the trigger pulse with 62 digitised channels for high and low sensitivity
read-out.
The trigger is given by the shaped sum signal of all the PMTs.

The scheme of the electronic chain is not available for any of the experiments apart from the ZEPLIN-II,
a study of the noise is not reported and it is not
clear how multiple events (due to background) are managed and discriminated with
respect to single events (the class to which WIMP-like events belong). In ZEPLIN-III 
it is specified that the $\chi^2$ goodness fit indicators within the position reconstruction of both
S1 and S2 are used to remove multiple-scatter events and in particular
double-Compton
events are discussed in more detail \cite{zeplas}.

Finally it would be interesting to read more comments on primary scintillation in gas (in addition to primary scintillation in liquid and
secondary scintillation in gas) which is undoubtedly weak but occurs just in front of the top PMTs, in experiments exploiting this configuration.

\section{Calibrations}\label{Cal}

Calibration is a fundamental aspect of any experiment, but 
is of particular importance and requires specific protocols
when considered in the framework of very low energy rare event searches. 
 The perfect recipe for a scintillator detector is given for instance by Birks \cite{Bir}
 requiring a good energy resolution, good linearity, uniform light collection,
 good light production to reach low threshold and suitable time response
 depending on the applications.

When scintillation light is considered in the field of direct Dark Matter
detection, many tests have to be added to the standard calibration procedures.
In the keV range
a precise characterisation of the detector is required, 
as well as studies of the detector response on the surface and on the bulk
with many low energy gamma and X-ray sources; moreover,
the possible effects of the surrounding materials have to be investigated, 
the energy threshold cannot be easily extrapolated,
and has to be evaluated with specific calibration system, 
the capability of discriminating noise at threshold has to be demonstrated and, 
internal lines identified. After these preliminary investigations,
calibration has to be performed  frequently, under operative conditions, in order to check the stability
of the:  energy threshold, electronics and PMT gains, energy scale, efficiencies, working conditions and in order to acquire the
statistics needed to define efficiencies and acceptance windows accurately.
In addition, there is the non-negligible aspect, that all the calibrations should be performed
in a sealed environment, free from Radon and under the same experimental
conditions as the production runs.

All these points may, on the surface appear trivial but are not for this field of physics. 
Very little information is given concerning calibration in the data published on double phase detectors.
XENON10, despite the large number of PMTs, only provides information concerning  the
S1 signal; no plots showing energy distributions for different sources on different scales are given or linearity
versus energy curves or energy resolution versus energy curves.
The situation is the same for ZEPLIN-II and ZEPLIN-III where a plot for S1 and S2 is given but only
one source is considered.
In WARP the plots and the energy resolution fit are only given 
in ref. \cite{arg1} where the measurement was performed
at zero electric fields. No information is given for the real operative working conditions.

The neutron calibrations of the detectors will be considered in the Section on the quenching factor and
are important in determining detector response and the energy threshold in terms of nuclear recoil energy.

XENON10  uses external gamma sources ($^{57}$Co, $^{137}$Cs) and 
gamma rays from metastable isotopes
($^{131m}$Xe and $^{129m}$Xe) produced by neutron activation of a xenon sample (450 g) 
and introduced into the detector after the production runs. These isotopes decay emitting 
164 keV and 236 keV gamma rays with half-lives
of 11.8 days and 8.9 days respectively. These photons can be used to check the response of the bulk of the detector
but not completely in the very low energy range. 
The authors explain that external, encapsulated sources can be inserted in the detector cavity
inside the lead and polyethylene shields \cite{xen2}.
$^{57}$Co and $^{137}$Cs sources are used from the external side and the external radial response
of the detector can be checked. The plots regarding these calibrations are given (see Fig. \ref{res}
at different radial values).  It can be seen that considering the large number of PMTs, these 
calibrations are mainly useful for the local external PMTs, but the area with a radius over 8 cm will be
later excluded by the fiducial radial cut. They state: {\it The light yield 
does not change substantially (less than 5\%) for radii r $<$ 9 cm} \cite{xen2}.

The WARP authors declare daily calibrations ``for long-term stability and linearity of the response'' and
the $\gamma$-ray sources used in ref. \cite{arg1} were: $^{57}$Co, $^{60}$Co, $^{137}$Cs placed outside 
the stainless steel chamber.
As mentioned at the beginning of Sect. 2 liquid argon has a lower absolute scintillation and ionisation efficiency
than liquid xenon and therefore major uncertainties exist for very low energy calibration procedures.

The authors of ZEPLIN-II report that calibrations are performed daily with an external $^{57}$Co source, 
placed between the detector and the scintillator veto by an automated mechanism. 
In different points the copper base was made thinner to allow penetration of 
122 keV and 136 keV $\gamma$-rays through to the bottom 1 cm layer of liquid xenon \cite{zep3}.

A preliminary calibration is available for ZEPLIN-III, performed during the commissioning tests,
with 31 internal $^{241}$Am at zero electric field and one external $^{57}$Co sources \cite{zep6}. During 
production runs, there are daily calibrations with a  $^{57}$Co source and a calibration with $^{137}$Cs source performed at the beginning and end of
data taking \cite{zeplas} (see Sect. \ref{ener}).

Energy resolution, energy threshold and uniformity of response are also strongly dependent on the design and on the
materials chosen for the detectors. Let us consider all  these aspects in detail.

\subsection{Energy resolution}\label{ener}

Decreasing scintillation when an electric field is applied in liquid noble gases  was originally
noted by ref. \cite{Ku78} and subsequently confirmed by other authors as already reported in Sect.\ref{Dat}.
  In the past, due to the anti-correlation between charge and scintillation
signals, detectors operating in scintillation and ionisation mode generally used the scintillation signal as the
event trigger. 
The energy resolution for scintillation for both liquid argon and xenon deteriorates in the presence of an electric field.
 A full systematics study of this behaviour for different noble liquids and particle excitation has not yet been published;
some investigations carried out on electrons, alpha particles and relativistic heavy ions can be found in refs. \cite{Apr04,Apr05,Craw,Craw2}.
This effect represents a limitation to the response of double phase detectors and the energy resolution obtained using them is worse than that obtained
 using other detectors used in Dark Matter direct investigations. This is especially true when considering that the energy calibrations
 near the claimed energy threshold are not available for double phase detectors. 

\begin{figure}
\begin{center}
\includegraphics[height=5.0cm]{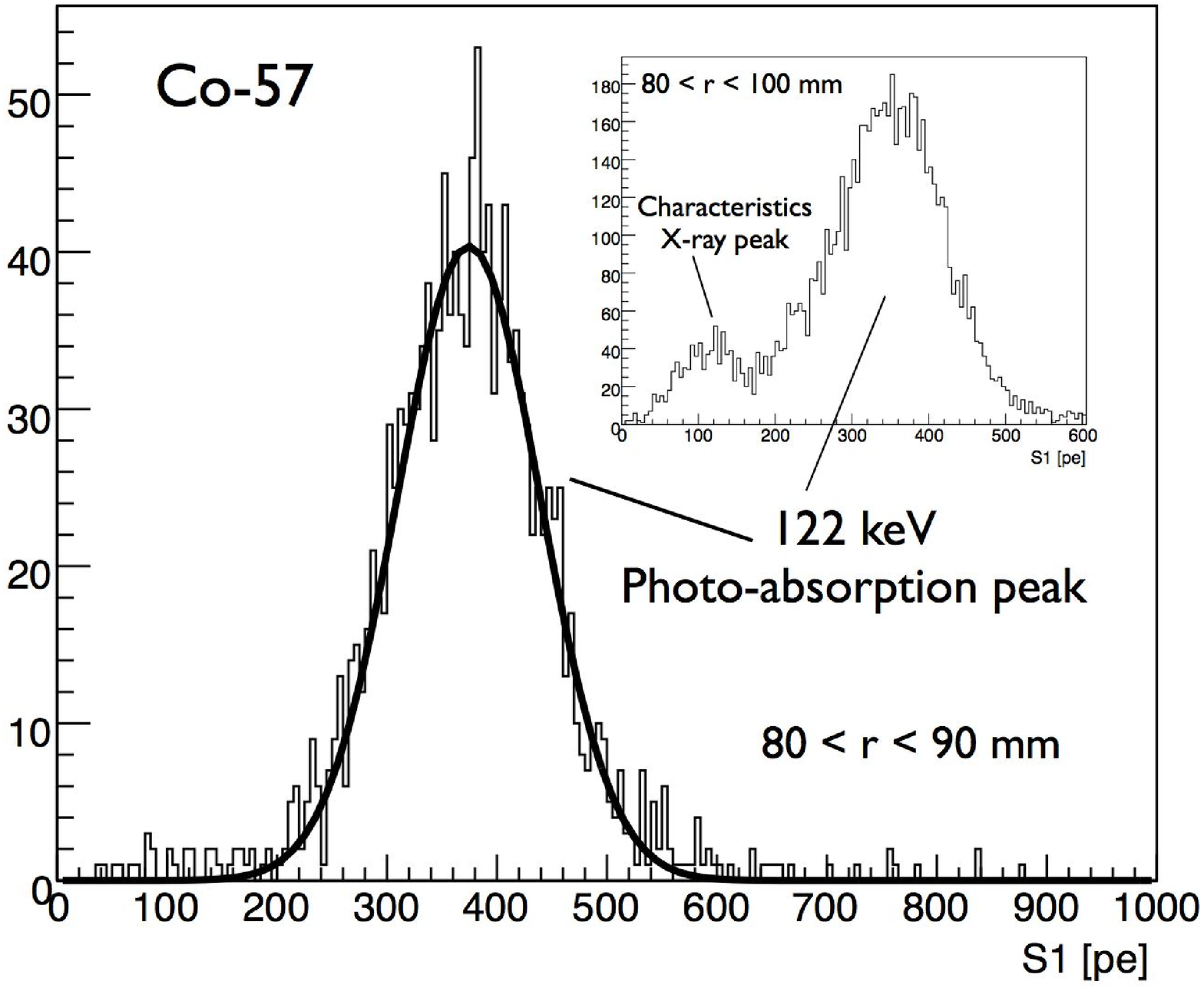}
\includegraphics[height=5.0cm]{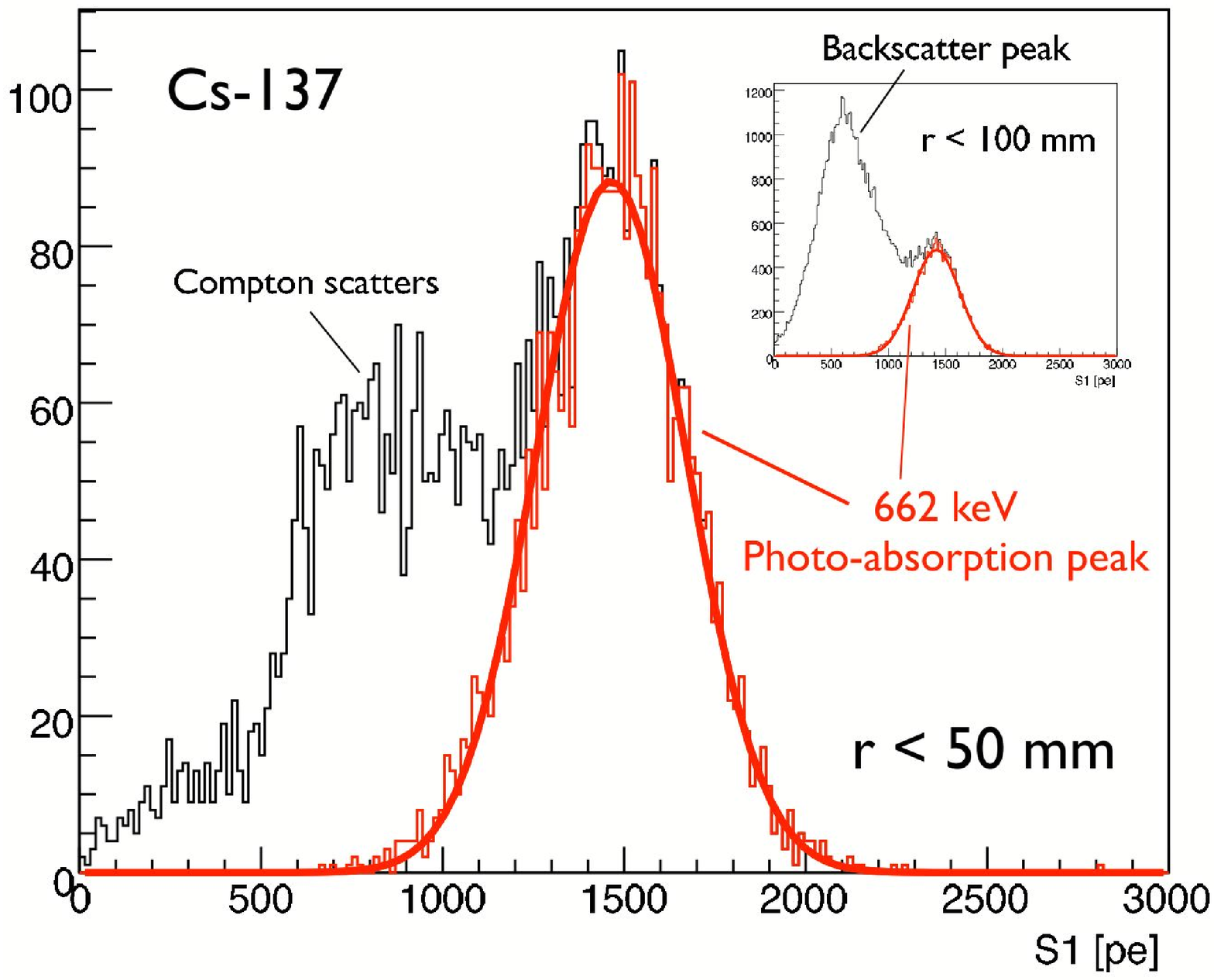}
\includegraphics[height=6.3cm]{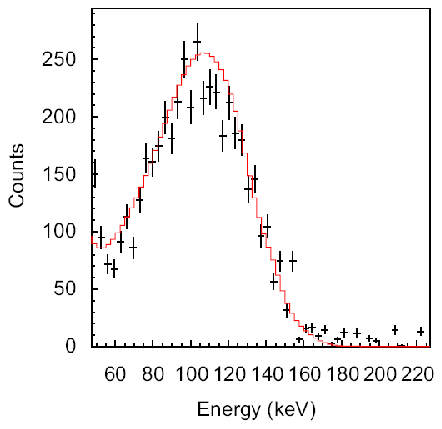}
\includegraphics[height=6.3cm]{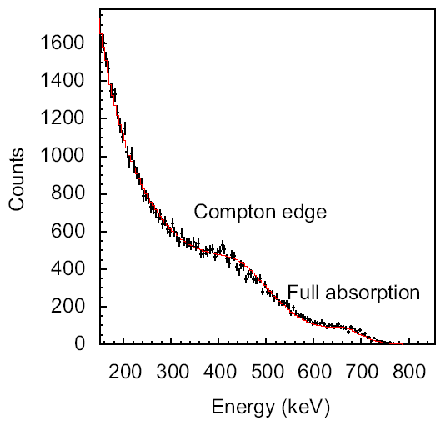}
\includegraphics[height=7.0cm]{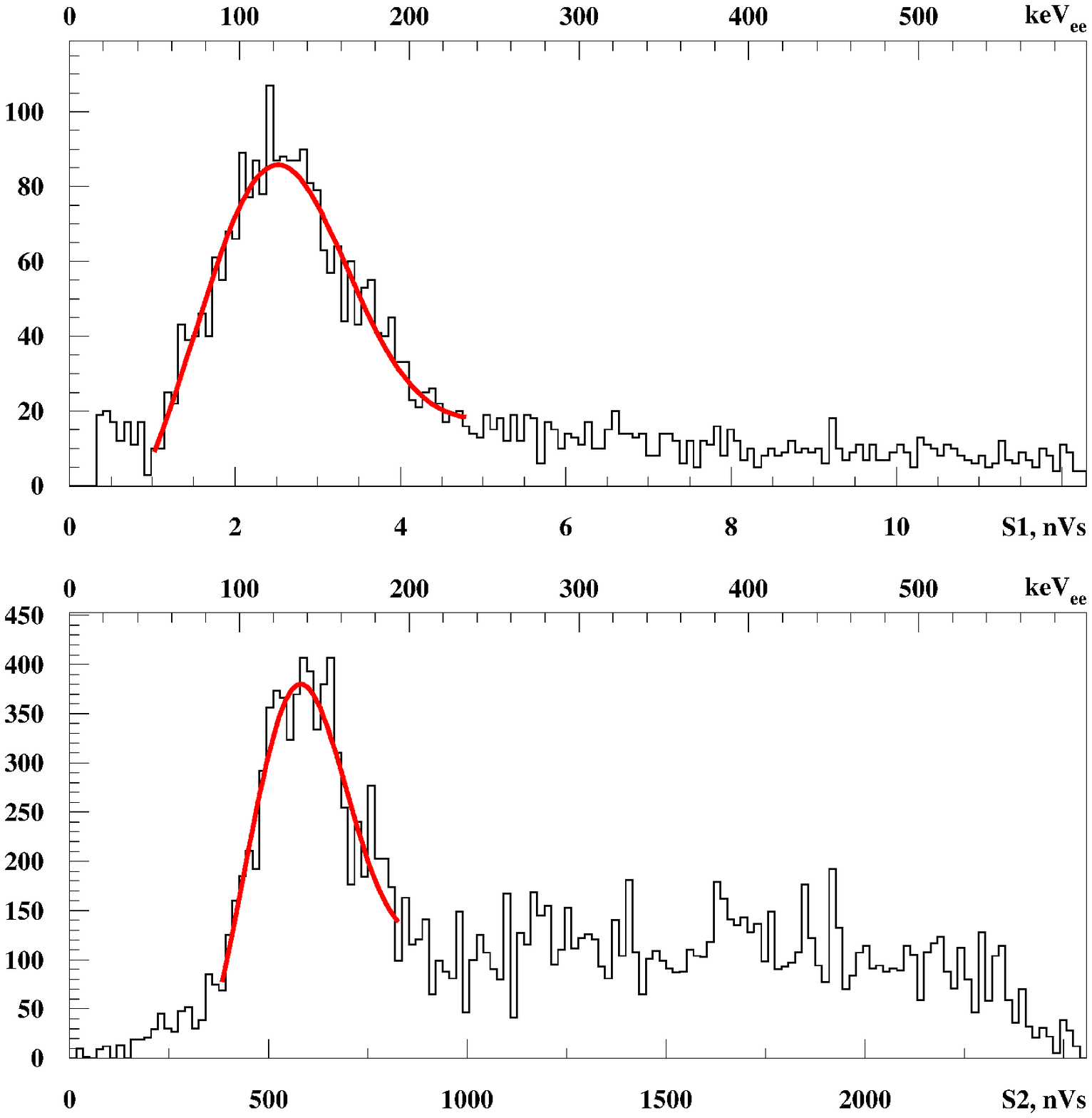}
\includegraphics[height=6.0cm]{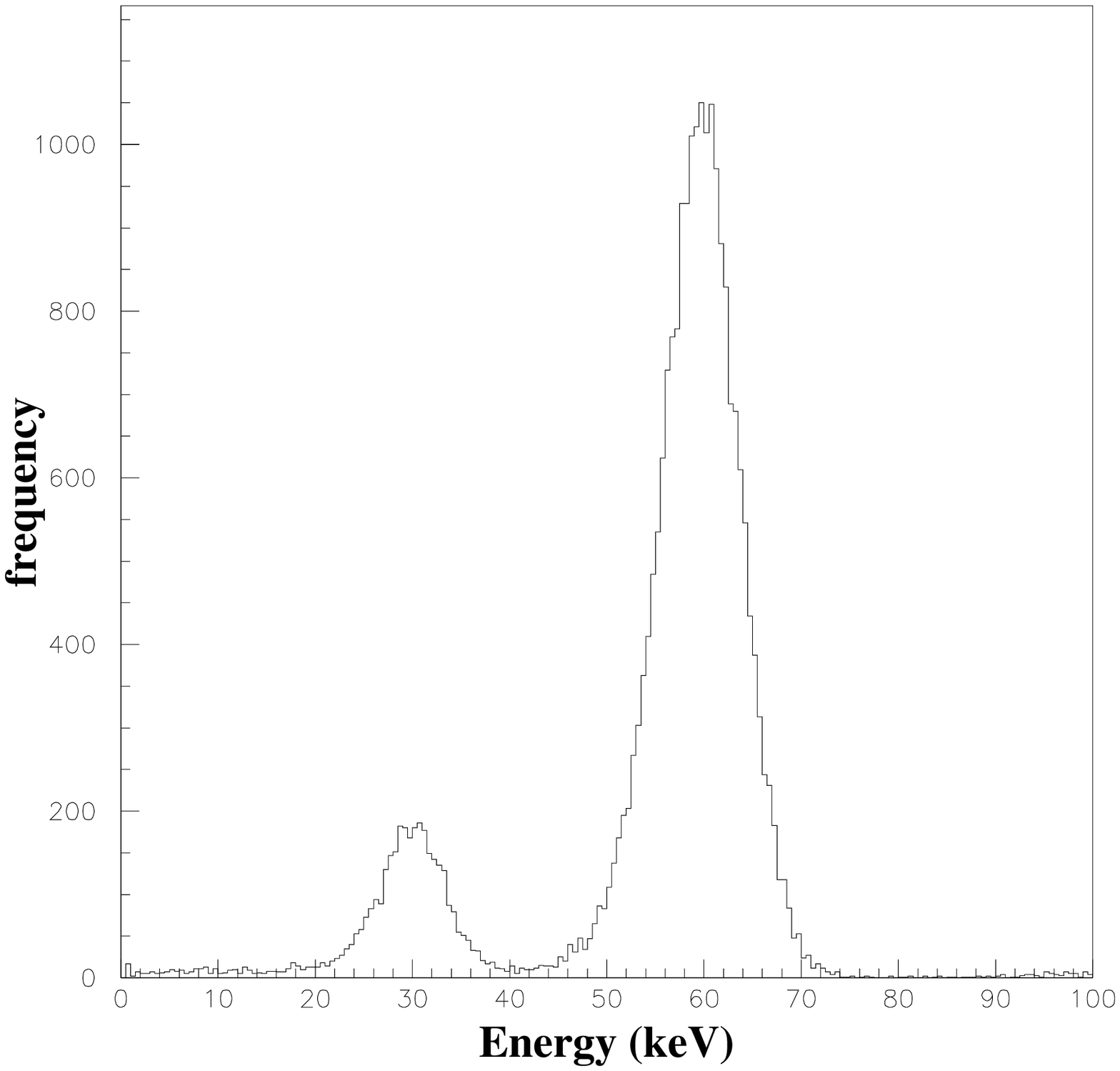}
\end{center}
\caption{Calibrations with $\gamma$ sources published by the different prototypes (figures taken from refs.
\cite{arg1,zep3,xen2}).
The typical spectrum obtained for $^{241}$Am with a NaI(Tl) crystal is also shown for comparison (bottom right).
Top: XENON10 - S1, $^{57}$Co and  $^{137}$Cs \cite{xen2}.
Middle: WARP - S1, $^{57}$Co and  $^{137}$Cs at zero field \cite{arg1}. 
Bottom left: ZEPLIN-II - S1 and S2, $^{57}$Co \cite{zep3}.
The figures clearly show the worse responses and energy resolutions 
of the Liquid Noble gases detectors with respect to NaI(Tl).}
\label{res}
\end{figure}

The energy spectra of gamma calibrations published by the different experiments are shown in Fig. \ref{res}.
 The typical spectrum obtained for $^{241}$Am with a NaI(Tl) crystal is also shown for comparison ($\sigma$/E= 6.8\% at 59.9 keV).

In XENON10, for a $^{57}$Co source and a radius 8 $<$ r $<$ 9 cm the obtained energy resolution 
is $\sigma$/E = 17\% at 122 keV.
In the case of the $^{137}$Cs source, the energy resolution is $\sigma$/E =14\% \cite{xen2} at 662 keV. 
It would be useful to know the energy resolution at the claimed energy 
threshold of 2 keV but this information is not given. This parameter
strongly influences the sensitivity claimed by the group.
 In fact, the crucial roles played by the energy threshold, the energy
 resolution at threshold, the energy resolution versus energy and other parameters (such as e.g. the escape velocity of the Galaxy)
 are well known in this field, especially for the sensitivity curves at low WIMP masses.
   In fact, owing to the quasi-exponential behaviour
 of the expected counting rate  (in their considered case) as a function of the detected energy, 
 even small changes in the energy threshold and resolution can affect the estimated sensitivity 
 up to orders of magnitude.

An energy resolution of $\sigma$/E = 13\% at 122 keV ($^{57}$Co) 
and $\sigma$/E = 6\% at 662 keV ($^{137}$Cs) \cite{arg1} was obtained at zero field by WARP
and linearity was reported between energy and primary 
scintillation S1.
Moreover, the position of the $^{57}$Co peak seems to be at $\simeq 105-110$ keV (see Fig. \ref{res}) well below
the expected one from  $^{57}$Co source photons.
The energy resolution $\sigma$(E) is parametrized:
$\sigma$(E) = $\sqrt{a^2_0 +a_1\cdot E + (a_2\cdot E)^2 }$ 
where a$_0$ =9.5 keV, a$_1$ = 1.2 keV and a$_2$= 0.04 \cite{arg1}. The value obtained for phe/keV is not reported.
One could argue $\simeq$0.5-1 phe/keV from the $^{57}$Co 
spectrum which could be worse
when the electric field was applied.

The energy resolution obtained in ZEPLIN-II, determined from the width of the $^{57}$Co 122/136 keV
$\gamma$ ray peak and other calibration lines (that were not reported), was  $\sigma$(E) = (1.80$\pm$0.04)$\cdot\sqrt{E}$ [keV], 
with E being the $\gamma$ ray energy in keV \cite{zep3}.

\begin{figure}
\begin{center}
\includegraphics[height=8.0cm]{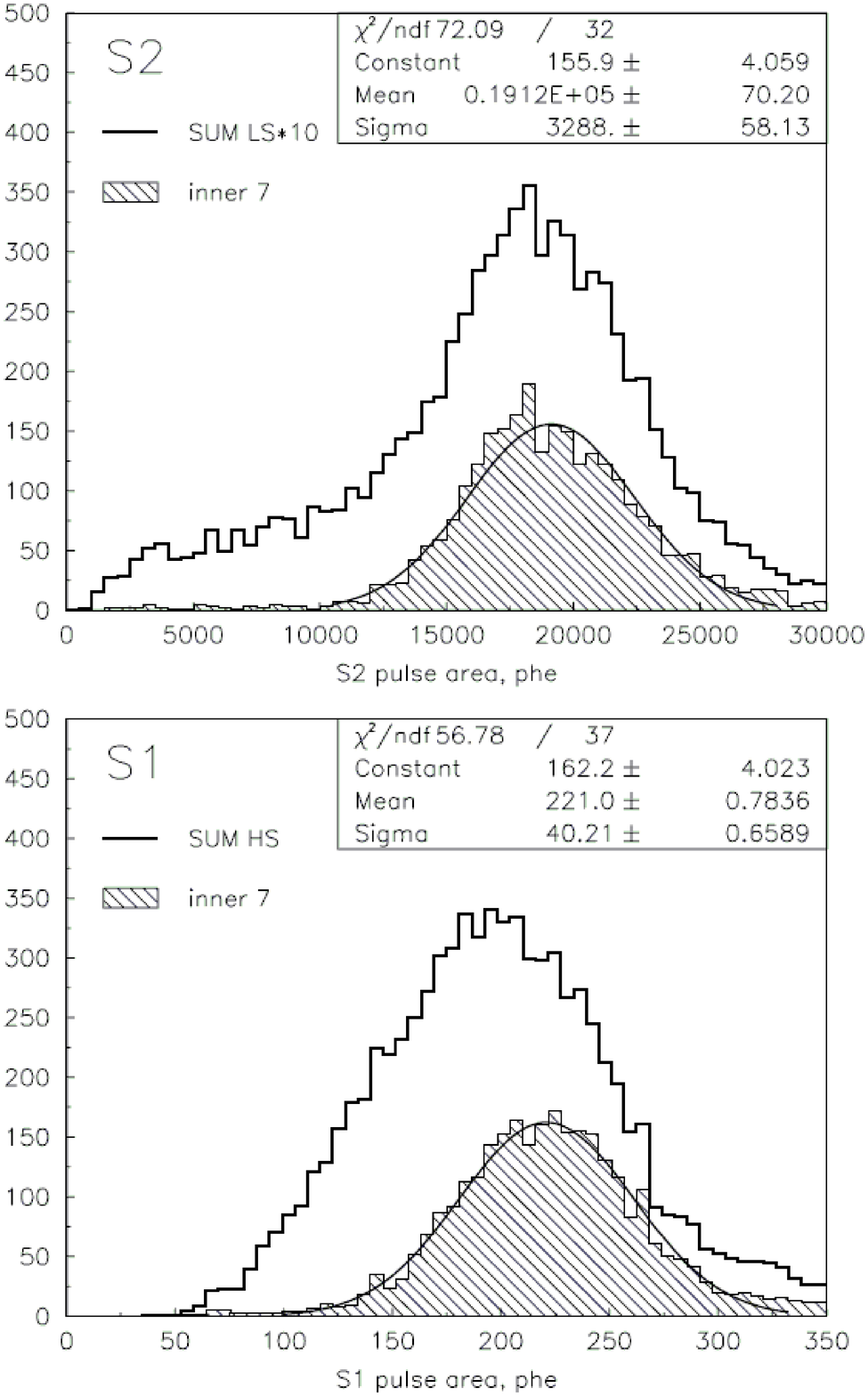}
\end{center}
\caption{ZEPLIN-III detector. Spectra for S2 (top) and S1 (bottom) from an uncollimated $^{57}$Co source above the instrument,
with a field of 3.0 kV/cm in the liquid. Even in the case of a ``collimated'' source, 
{\it with peak signals in one of the inner 7 PMTs}, 
a light yield of 1.8 phe/keV ($\sigma$/E$\simeq$ 18\%) is quoted \cite{zep6}.
Figure taken from ref. \cite{zep6}.}
\label{res1}
\end{figure}

A preliminary calibration is available for ZEPLIN-III, performed during the commissioning tests, by substituting 
the electrodes with a copper plate above the PMT array holding 31 $^{241}$Am sources (one per PMT). 
By using the MCA single photoelectron spectrum of one PMT they report $\simeq$12 phe/keV for the liquid phase without
 applying an electric field and $\sigma$/E = 6.4\%. 
The value of phe/keV is impressive considering that  for ideal Poissonian statistical behaviour the same energy
resolution corresponds to 4 phe/keV; this resolution can generally be obtained with an external $^{241}$Am source  
using an NaI(Tl)  if  the detector has
around 5--7 phe/keV. As specified in the paper \cite{zep6} the source is a few mm from the PMT.
An improved light collection (up to 17 phe/keV) for the same configuration is reported when the liquid
level is between the source and the PMT window,
and the obtained two-phase resolution is $\sigma$/E$\simeq$ 5.5\%
at 59.5 keV. This clearly does not indicate the values of the energy resolution if the
sources were placed in different position within the total volume.
A few mm of liquid or (liquid + gas) 
does not help, indeed the operative conditions correspond to a deeper liquid gap.
In fact, when the liquid fills the nominal depth, a calibration is performed with an uncollimated $^{57}$Co source
above the detector without an electric field. This time it is not specified if the result is for a single PMT, but in any case the energy
resolution $\sigma$/E$\simeq$ 10.6\% is reported for the fiducial volume. This value is comparable with a NaI(Tl) response,
but the authors declare 5 phe/keV (a value that in any case better than the one they obtained 
using Monte Carlo simulation). 
 When the electric field is applied (3 kV/cm) energy resolution obtained
is shown in Fig. \ref{res1} for the $^{57}$Co spectrum and all the PMTs ($\sigma$/E$\simeq$ 25\%). 
At the end even in the case of a ``collimated'' source, 
{\it with peak signals in one of the inner 7 PMTs}, a light yield of 1.8 phe/keV ($\sigma$/E$\simeq$ 18\%) 
is reported \cite{zep6}.
During production runs the light yield of 1.8 phe/keV was also confirmed at the working drift field (3.9 kV/cm) \cite{zeplas}.

\begin{figure}
\begin{center}
\includegraphics[height=5.0cm]{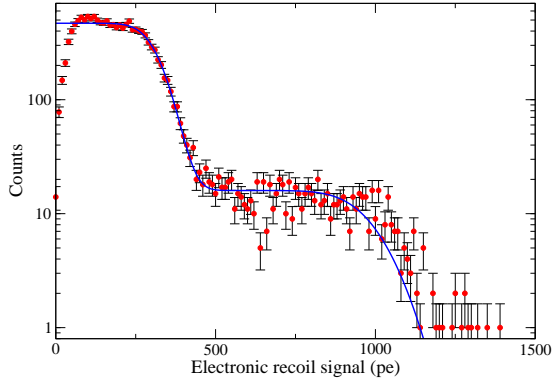}
\end{center}
\caption{Energy spectrum of $^{22}$Na calibration data,
reporting a signal yield of $0.93 \pm 0.07$ phe/keV \cite{Mac2}. 
Figure taken from ref. \cite{Mac2}.}
\label{res2}
\end{figure}

We know little about neon scintillation, a lower light yield with respect to that of argon is expected, which in its turn is lower than that of xenon.
In fact poor energy resolution is obtained with liquid neon as can be seen from Fig. \ref{res2} obtained with an external $^{22}$Na source,
reporting 0.93$\pm$0.07 phe/keV \cite{Mac2}. The data are from two PMTs requiring that the integrated signal from one PMT be within a factor of four of the other.
On four days of measurement the yield varies by 7\%. Notwithstanding the many technical problems that have yet to be tackled, ton and multi-ton detectors have already been proposed.

Studies performed on scintillation luminescence in liquid phase have shown that energy resolution is dependent
on liquid density (i.e. on temperature) and improvements in energy resolution have been obtained both for ionisation
 and scintillation by increasing the temperature 
(see for instance ref. \cite{Obo,Kir}). 
 It has also been suggested that the improved energy resolution obtained for liquid argon with respect to
liquid xenon is due to the smaller density of matter \cite{Obo}.

A density dependence is also present in the case of high pressure xenon gas (high density) \cite{Bol,Koba} for which 
the scintillation luminescence at zero field shows a decreasing  behaviour when density increases.

In conclusion, this Section has described the present status of energy resolution
 in double phase liquid noble gas detectors used for Dark Matter direct investigation, highlighting reliance on the running parameters 
 and outlining problems and present limitations.

\subsection{Light collection}\label{Uni}

When working with scintillators, a general principle has to be satisfied: identical events have to produce identical
amount of light, i.e. the amount of light produced by an event must be independent 
of the point in the sensitive volume where the event happens.
The uniformity in the response of a scintillator can be conditioned by many factors:
the attenuation length, purity and in the case under study here, 
the design of the detector and electric fields, etc..

A lack of uniformity in light collection was already reported in 
ZEPLIN-I, where a position dependent light collection was found 
and a geometrical correction was performed using ``rebinning matrix'' calculated via
a Monte Carlo code. The authors explain that $^{57}$Co
(122 keV) and $^{137}$Cs (660 keV) sources placed under
the xenon vessel were used. In the case of the $^{137}$Cs source a measured light yield 
 25\% lower than the $^{57}$Co source was found.
For the authors the reason is that:
{\it  due
to a position-dependent light collection, the
122 keV gammas penetrating only a few mm into
the bottom of the xenon vo\-lu\-me, while the
662 keV gammas deposited energy (by Compton
scattering and absorption) throughout the chamber
volume} \cite{zep1}. 
In order to take this into account, the position-dependent
light collection is evaluated by a simulation code.
At the end this ``rebinning matrix'' is used  together with an ``energy resolution matrix''  
for the evaluation of the Dark Matter cross section limits \cite{zep1}.

One could argue that this is only an intrinsic limit of ZEPLIN-I due to its
 specific work conditions
 and in fact in ZEPLIN-II it is declared that 
the observed
light collection throughout the active volume was uniform to within 3\% as derived from the study
of the primary scintillation signals for $\alpha$ particles \cite{zep3}.
 Notwithstanding this statement non-uniformity is still present because
the same procedure is also applied in ZEPLIN-II with the position dependent
 correction and is justified as an effect of the energy
  resolution:
{\it  This has the effect of mixing the events
between energy bins, which can at the final stage of analysis be accounted for
by applying a compensating rebinning matrix to the energy-binned spectral
terms, as shown in detail in [7]} (i.e. the procedure applied in ZEPLIN-I is cited) \cite{zep1}. 

And again, at the end of the paper, when exclusion plot limits are calculated
under a single set of assumptions and without including experimental uncertainties,
the authors write a list of instructions and the first one is:
{\it Apply an energy resolution correction as described in greater detail in
a previous paper [7], by numerically applying the resolution rebinning
matrix to the vector of binned spectral terms} \cite{zep3}.

Instead of using a Monte Carlo procedure as in ZEPLIN, the
XENON10 carried out position correcting \cite{Kni,xen6,fio,Ober} to S1 and S2 signals 
with the maps obtained from activated Xe: (in the first version of the paper \cite{xen1})
{\it The S1 and
S2 response from the $^{131m}$Xe 164 keV gamma rays,
which interact uniformly within the detector, 
were used to correct the position dependence of the two signals}. While in the final version
they say: {\it The S1 and
S2 response from the $^{131m}$Xe 164 keV gamma rays,
which interact uniformly within the detector, 
were used to correct   $\pm$ 20\% variations 
of the signals due to the position dependence of the light 
collection efficiency.}
They also write that after position-dependent
corrections, from S1 for 122 keV gamma rays, 
a volume-averaged light yield, Ly, of
3.0 $\pm$0.1(syst)$\pm$0.1(stat) phe/keVee at the drift field of
0.73 kV/cm is derived \cite{xen1}.
More recently, position dependent corrections have been reported \cite{Ober} in
$x,y,z$ for S1 signal and in $x,y$ for S2.

A trace of non-uniformity in light collection is also found 
in WARP papers, where for the formula used to evaluate
energy resolution at zero field (see Sect. \ref{ener})
it is explicitly stated that
the term a$_2$ takes
into account the effects from non-uniform light collection \cite{arg1}. 
In ref. \cite{arg2} the
exclusion plot calculation is written about again:
{\it The energy resolution has been taken into account considering the statistical fluctuations in the number
of collected photoelectrons and the geometrical effects due to non-uniform light collection ($\sim$ 5\%) as deduced from
calibrations} \cite{arg2}.
However there is no possibility of checking this value from
their paper given that the information needed from the gamma calibrations is not provided.

The ZEPLIN-III group referring to Fig. \ref{res1}
writes that the broadening 
on the low side for the uncollimated spectra is 
 due to light collection variations towards the edge of the detector so:
{\it Most of these fall outside the fiducial
volume and the remainder can be corrected using 3-D position reconstruction information} \cite{zep5}.
And again in ref.\cite{zeplas} a procedure of position correction to S1 and S2 responses is confirmed and described.

The lack of uniformity in these detectors (due to the different reasons mentioned at the 
beginning of this section, but also arising from intrinsic limits) means that position-dependent corrections must be made 
 inferring on the event identification and on the definition of the fiducial volume. These procedures are 
 highly questionable, uncertain and critical in the Dark Matter detection due to the very low energy nature 
 of the involved processes.
In fact, it is well known that similar corrections can introduce non-negligible systematics
 which can affect the sensitivity of detector. These problems have not yet been tackled.

Finally all experimental parameters should be reported. Different experiments did not give 
data pertaining to several parameters including light collection efficiencies for proportional and direct light.

\subsection{More on light yield stability}

\begin{figure}
\begin{center}
\includegraphics[height=6.5cm]{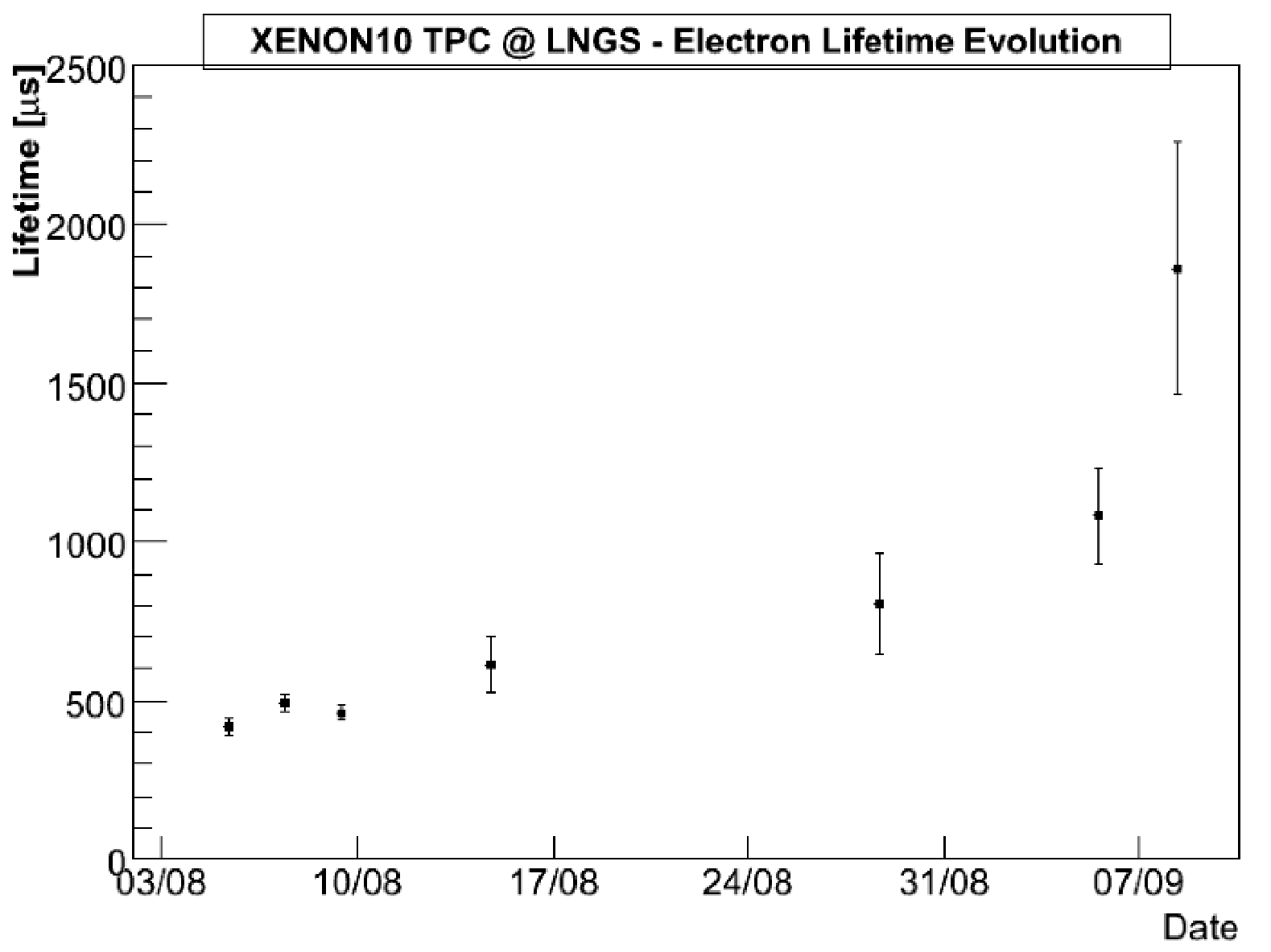}
\includegraphics[height=5.5cm]{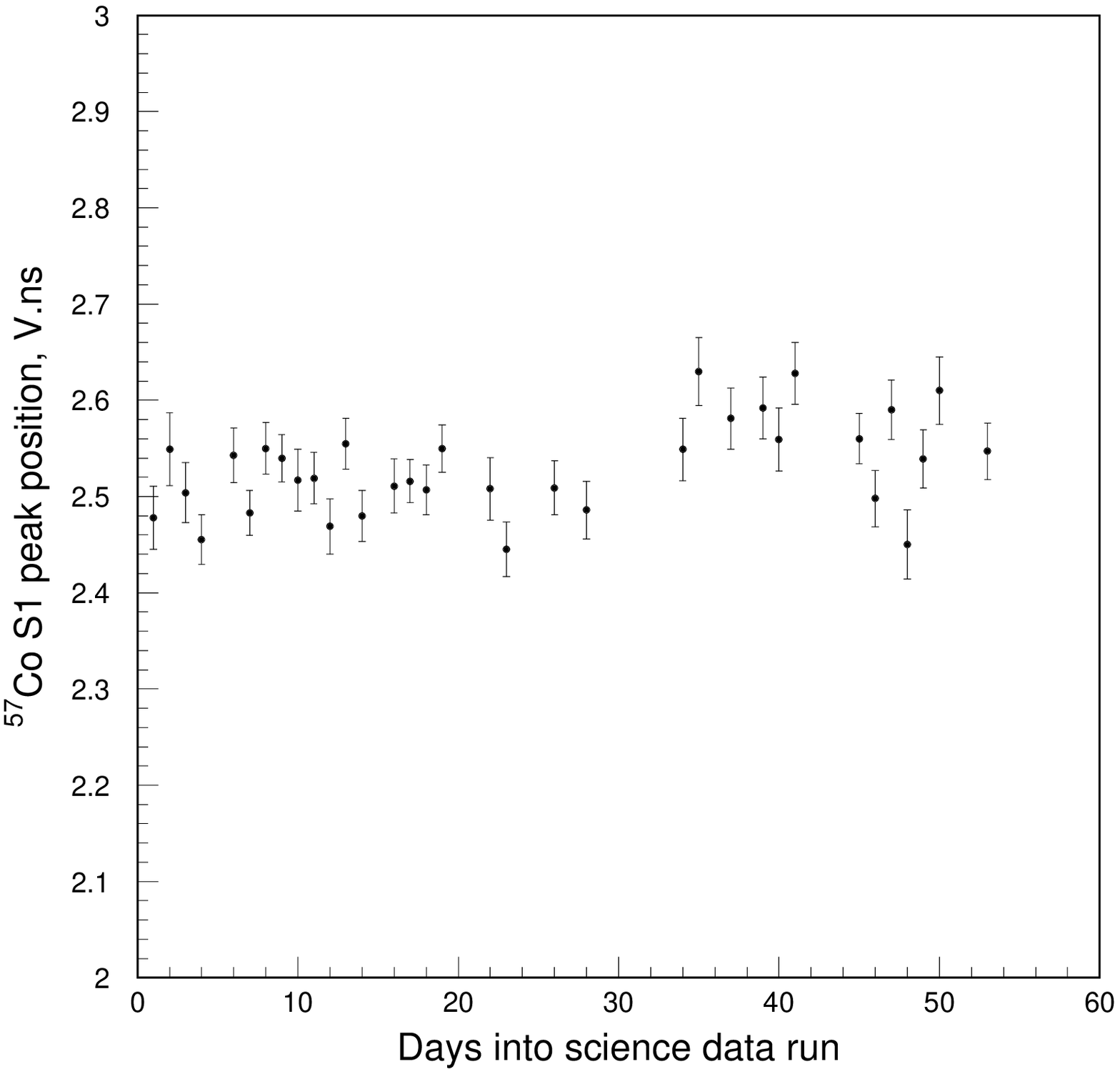}
\includegraphics[height=5.0cm]{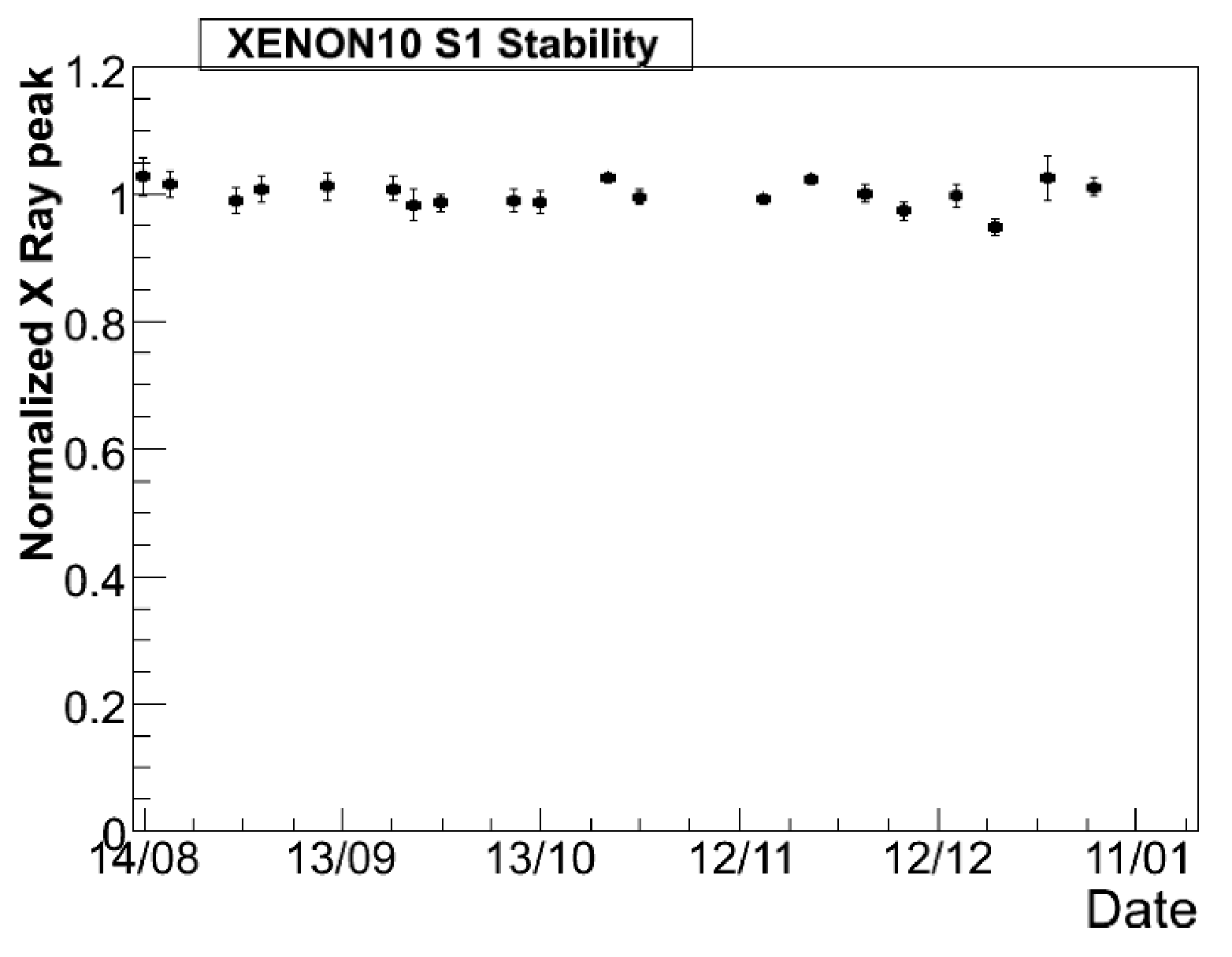}
\end{center}
\caption{Top: XENON10, lifetime evolution \cite{xen7}. 
Bottom: source peak position versus time; left - ZEPLIN-II $^{57}$Co \cite{zep3}, 
right - XENON10 X-ray 37 keV \cite{xen7}.}
\label{xsta}
\end{figure}

Both XENON10 and ZEPLIN-II show the plots with the behaviour of the X-ray 37 keV, $^{137}$Cs \cite{xen7} and
the $^{57}$Co \cite{zep3} peak respectively as a function of the time in order to check light
yield stability (see Fig. \ref{xsta}). 

XENON10 also illustrated \cite{xen7} (see Fig. \ref{xsta}) the electron lifetime evolution that 
increases with time. The S2 signal depends on the charge collection, 
so probably S2/S1 changes with time. It would be useful to see the energy resolution evolution 
with time in order to understand how critical working conditions are 
for the stability of the acceptance windows.

\subsection{ Linearity of light yield and energy threshold}\label{Nl}

The XENON10 paper \cite{xen2} reports for the $^{57}$Co source an  
average light yield for 122 keV gamma rays of 374 photo-electrons (phe) (i.e. 3.1 phe/keV) while
for the $^{137}$Cs (662 keV photo-absorption peak) an average light 
yield of 1464 phe (i.e. 2.2 phe/keV).  This means a non-linearity of the energy scale,
so it is difficult to understand how the energy threshold can be fixed in these conditions without specific calibration
at very low energy and 
what the actual value of phe/keV is at very
low energy and the corresponding efficiency and the keV energy scale.

At present, the light yield measured by XENON100 is even worse than XENON10
as the authors themselves indicate\cite{xcen}: {\it Figure 4 shows the current S1 light spectrum of $^{137}$Cs 662 keV gamma-rays,
measured in XENON100, showing a light yield of 2.7 pe/keVee (zero field), limited by the
presence of impurities, mostly H$_2$O}.

ZEPLIN-II only reports one gamma source (they also mention a $^{60}$Co source
but they do not give any plots) writing at the end that the average photoelectron yield for the PMTs
was 1.10$\pm$0.04 phe/keV with the electric drift field set to zero, and
0.55$\pm$0.02 phe/keV with the electric field at 1 kV/cm (the operating value) \cite{zep3}.

For WARP the value of phe/keV is not given in the paper \cite{arg2}. In a previous configuration with the 2.3 l detector
viewed by one 8inch PMT they reported 2.35 phe/keV for a drift field of 1 kV/cm and a $^{109}$Cd source \cite{arg3}.

\section{Luminescence yield and time response}

The linear energy transfer (LET) dependence of the luminescence yield and the
time response in liquid argon and xenon, were investigated in depth for electron, $\alpha$-particle and fission-fragment
 excitation \cite{Hit,Kub,Car,Ket,Sue,Him}. The time response
  is characterised by
two decay constants, corresponding to the life time of the singlet state and of the triplet state of the excited molecule.
Some results are summarised for liquid argon, xenon and neon in Table \ref{tab2}.
\begin{table}[htbp]
\caption{Decay times for the singlet $\tau_s$ and  triplet $\tau_t$ components for liquid argon, liquid xenon and liquid neon. The intensity
                ratios of the two components are also shown. f.f. stands for fission fragments.}
\begin{center}
\begin{tabular}{ccccccc}
\hline
   Liquid         & Particle               & $\tau_s$ (ns)     & $\tau_t$ (ns)          &   $I_s$/I$_t$ & Ref.       \\
\hline
  argon          & electron              & 6$\pm$2            &    1590$\pm$100  &          0.3        &   \cite{Hit}\\
      ~              & ~                           & 6                        &     1200                    &                         &   \cite{Him} \\
     ~              & ~                           & 4.6                       &     1540                    &          0.26      &   \cite{Car} \\
     ~              & ~                           & ~                           &     1110  $\pm$50  &        ~              &   \cite{Sue} \\
     ~              & ~                           & 4.18$\pm$0.2    &     1000  $\pm$95  &        ~              &   \cite{Ket} \\   
      ~             & ~                           & 6.3$\pm$0.2      &     1020$\pm$60     &          0.083    &   \cite{Kub} E=0 kV/cm\\
      ~             & ~                           & 5.0$\pm$0.2      &     860$\pm$30        &          0.045    &   \cite{Kub} E=6 kV/cm\\
       ~            & $\alpha$             &  7.1$\pm$1.0     &   1660$\pm$100      &     1.3              &   \cite{Hit}\\
       ~              & ~                         & 4.4                       &     1100                      &        3.3           &   \cite{Car} \\
      ~             & ~                           & $\sim$5               &     1200$\pm$100   &                        &   \cite{Kub} E=0 kV/cm\\              
          ~         & f.f                          & 6.8$\pm$1.0      &  1550$\pm$100        &          3             &   \cite{Hit}  \\
 \hline
   xenon      & electron               &   ~                         &                 45               &         ~               &  \cite{Hit} \\
     ~              & ~                           &   ~                         &             32$\pm$2      &         ~               &  \cite{Ket}  \\
      ~             & ~                           &   ~                          &            34$\pm$2      &         ~                 &   \cite{Kub} E=0 kV/cm\\   
    ~               & ~                            &  2.2$\pm$0.3     &     27$\pm$1             &          0.05          &   \cite{Kub} E=4 kV/cm\\  
    ~               & $\alpha$             &  4.3$\pm$0.6     &     22$\pm$1.5           & 0.45$\pm$0.07 & \cite{Hit} \\
      ~             & ~                           &   4                         &     27                            &  1.1                 &    \cite{Kub} \\
      ~             & ~                           &   3                          &    22                            &   3.4                & \cite{Kub} \\
     ~              &  f.f                         & 4.3$\pm$0.5       &     21$\pm$2              &  1.6$\pm$0.2   & \cite{Hit}\\
 \hline
  neon          &  electron           &  12.8                        & 2400                          &     2.8               & \cite{Nik}   \\
       ~              &                          &    --                           & 2900                          &                          & \cite{Sue}  \\
      ~              &                           &     --                          & 3900$\pm$500       &      ~                  & \cite{Mich} \\
     ~               & $^*$                  & 18.2$\pm$0.2       & 14900$\pm$300      &    ~                      & \cite{kec,Mac2} \\
\hline
\end{tabular}
\end{center}
$^*$ Obtained with electron and nuclear recoil. The authors attribute this increased value
 of the triplet component to the increased reached purity.
\label{tab2}
\end{table}
It is important to remember that the presence of impurities can
affect these values as can different temperature \cite{Hit}. Nitrogen contamination at ppm levels, for example, \cite{Nor1,Him}
significantly reduces the slow decay time in liquid argon, by up to two orders of magnitude.
Greater quantities around 350 ppm can result in even two exponential
decay constants in the slow component \cite{Him}. In refs. \cite{Kub1,Hit} a third intermediate component is reported
with a decay time around 20-40 ns the intensity of which is around 10-20\% of the total intensity, but no explanation is given.   
 In WARP \cite{arg2} a reduction in the slow decay time with respect to the values of ref. \cite{Hit}
is found and ascribed to a nitrogen contamination. The authors also mention that this also modifies the ratio of electron 
and recoil response. It should be underlined that the values reported in ref. \cite{Hit} were measured in the absence of electric field.
When an electric field is applied as in the case of ref. \cite{Kub} a decrease in the triplet component is observed for electron excitation. The effect of the electric field
in decreasing the triplet component is even more drastic in liquid xenon. 

The electric field also produces a decrease in the luminescence intensity in liquid argon and xenon excited by electrons. This decrease in the intensity 
is about 64\% at 6 kV/cm \cite{Kub} for liquid argon and 70\% at 4 kV/cm \cite{Kub} in liquid xenon relative to intensities without an applied electric field.

With liquid xenon the values obtained for the decay times are in agreement for  $\alpha$-particle excitation \cite{Hit,Kub}. 
Under electron excitation there is a non exponential contribution from recombination that is not present when an electric field is applied.
The triplet decay times are different in the papers \cite{Hit,Kub}.

The intensity ratios $I_s$/I$_t$ of the singlet states to the triplet states are reported in Table \ref{tab2} for
 electron, $\alpha$-particle and fission-fragments respectively for liquid argon and liquid xenon. 
The intensity ratios $I_s$/I$_t$ of the singlet states to the triplet states for argon and xenon vary in different papers \cite{Hit,Kub,Car,Ket,Sue}
for the reasons given above
 but they show the same trend, i.e. an increased value when the deposited energy density  is increased.

In WARP  the S1 and S2 signals are distinguished due to the different
time shape. For the primary scintillation S1 signal, the time shape is expected to be different in the case of a heavy particle with
respect to $\alpha$ and $\gamma$ radiation. WARP gives an example of the typical behaviour of the two cases for S1 signal \cite{arg2}. 
WARP introduces a pulse shape fast discrimination parameter F = I$_s$/(I$_s$+I$_t$). The minimum number of photoelectrons
(I$_s$+I$_t$) in the S1 signal is reported greater than 50 corresponding to an energy loss for argon recoils of 40 keV.
They state (?): {\it To evidence the necessity of a sufficient light yield in 
order to perfect the pulse shape separation we have artificially blanked the signals of four out of the seven photomultipliers.
With only photo-multipliers left, a very 
substantial worsening of the distributions of parameter F 
is observed, as expected} \cite{arg2}. The values found for F have nearly Gaussian distribution centered 
at F= 0.31 for electrons and F=0.75 for argon from 
neutron recoils, that corresponds to I$_s$/I$_t$= 0.45 for electrons and I$_s$/I$_t$= 3 for argon recoil
(see Table \ref{tab2} for a comparison with values available in literature).

In conclusion, this section has addressed the main aspects regarding the luminescence yield and 
the time response for liquid noble gases, pointing out their dependences on the impurities, 
on the electric field and on the nature of the particle.

 \section{Rejection procedures}\label{Rej}

In the framework of Dark Matter investigations, the main aim of this low energy application of double phase
detectors is the exclusive selection of nuclear recoils with respect to all the other
components of the counting rate, exploiting the different expected behaviour of S1 and S2 signals. 
In practice, for a number of reasons, many cuts are applied to the collected data
to differing extents by all the experiments. Let us consider the main aspects of this.

For all the experiments considered, the selection of the fiducial
volume requires a significant reduction of the sensitive volume, performed on the drift time
and in some cases by a further radial cut.

The reason is explained in more detail in the ZEPLIN-II paper \cite{zep3}.
End-range  $\alpha$ particles from boundary walls, from 
cathode and field grids can release energy down to keV range and they can mimic nuclear recoils.
They try rejecting these events with a timing cut. In addition nuclear recoils are produced by the 
 Radon progeny electrostatically attracted to the side walls. These events are
 dangerous because can mimic low energy xenon recoils. So a radial cut is applied to remove 
 these events and also low energy electron recoils from the walls \cite{zep3}.

XENON10 explains that for analyses the fiducial volume is 
delimited within 15 to 65 $\mu$s (about 9.3 cm in 
z-coordinate of the total 15 cm) and from a radius less than 8 cm (out of 10 cm) in xy-coordinates.
In XENON10 the definition of the fiducial volume is done also because:
{\it The
cut in Z also removes many anomalous events due to the
LXe around the bottom PMTs, where they happen more
frequently compared to the top part of the detector} \cite{xen1}.

WARP states that the fiducial volume cut is performed
on the drift time, explaining that this cut excludes events near the liquid surface
($\leq$2 cm) {\it where the primary pulse is not sufficiently
separated from the secondary for an appropriate event reconstruction} \cite{arg2} and
events due to the contaminations near the cathode region ($\leq$ 0.7 cm).

\begin{figure}
\begin{center}
\includegraphics[height=8.cm]{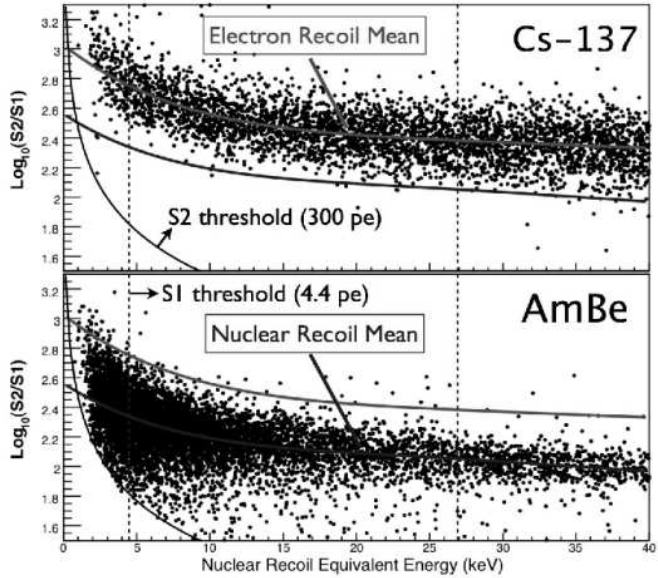}
\end{center}
\caption{XENON10: Log$_{10}$(S2/S1) as a function of energy for electrons
(top) and nuclear (bottom) recoils from calibration data.
Figure taken from ref. \cite{xen1}.}
\label{dis}
\end{figure}

The main parameter considered for the discrimination is the ratio S2/S1. 
This quantity or its Log$_{10}$ is plotted versus the nuclear recoil energy 
as in the case of ZEPLIN-II and XENON10, or versus the F parameter
in the case of WARP.
Typical behaviour is given for XENON10 in Fig. \ref{dis}, in the case of gamma and neutron sources.

In particular referring to Fig. \ref{dis}, that is Fig. 1 in the XENON10 paper, they note that:
{\it The separation of
the mean Log$_{10}$(S2/S1) values between electron and nuclear
 recoils increases at lower energy. In addition, the
width of the electron recoil band is also smaller at lower
energy} \cite{xen1}. 
The authors assume that 
the combination of these two effects gives a better selection of the electromagnetic component 
of the counting rate at lower energy, for which they say
efficiency values down to 99.9\% can be reached.
They suggest that this effect could be due to the different ionisation density
and track structure of the events \cite{xen1} and
confirm that this effect was also noted in previous prototypes \cite{xen5}.

The two distributions overlap both in XENON10 and ZEPLIN-II. For liquid xenon 
 the discrimination is even worse with respect to bolometer-ionisation technology.
Being present in both the detectors it is not clear if
it is a limitation of the present application of the technique or if it is an intrinsic limit. ZEPLIN-II 
also states that discrimination improves at low energy but this is not supported by general 
experience on scintillators. Is it a possible effect of the many cuts?

In WARP the authors only mention three different cuts applied to data: a cut on
the drift time, 10 $\mu$s $\le$ T$_d$ $\le$ 35 $\mu$s, corresponding to the fiducial
volume cut; a cut on the ratio S2/S1 and a cut on the parameter of pulse shape F.

\begin{figure}
\begin{center}
\includegraphics[height=10.0cm]{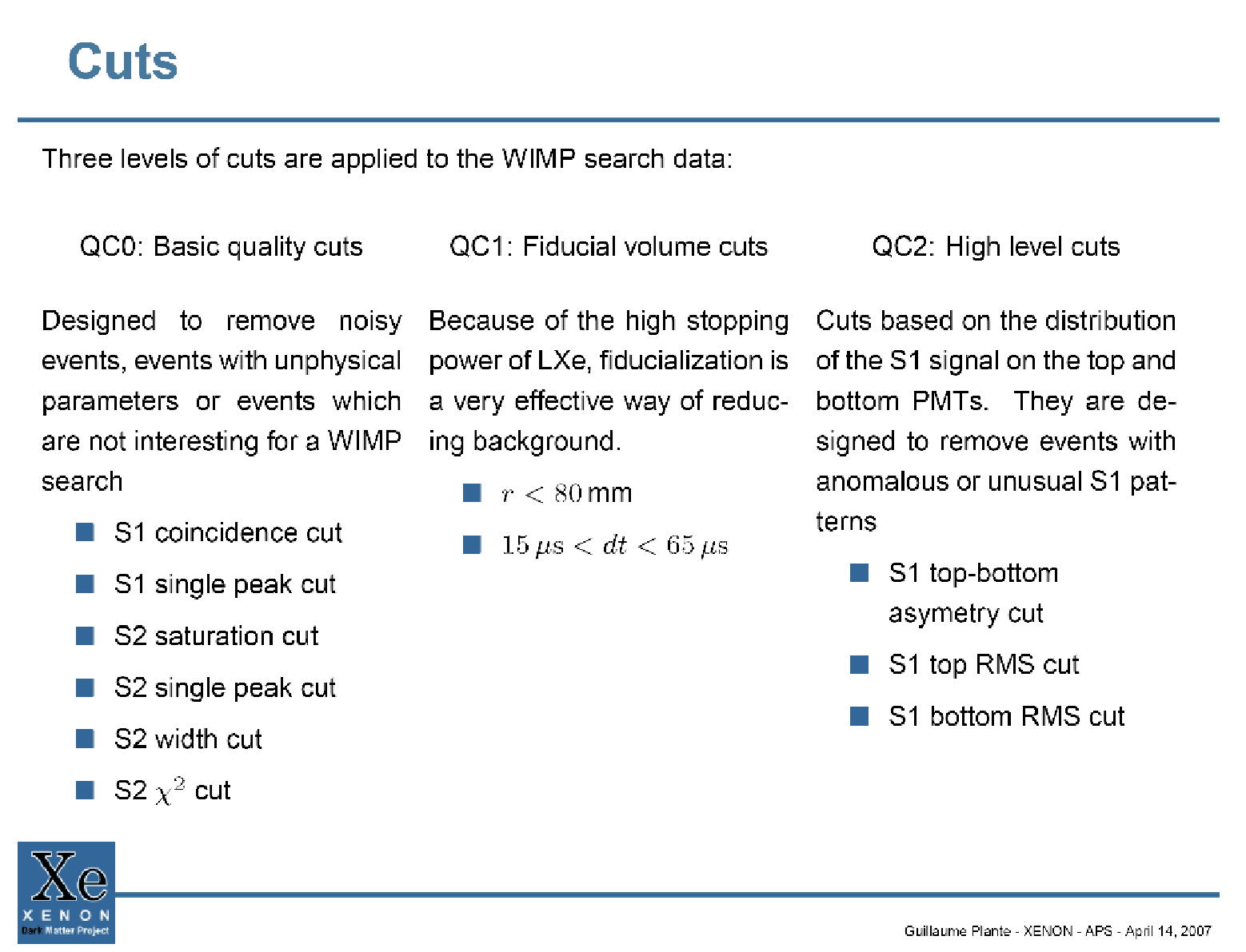}
\end{center}
\caption{List of applied cuts in XENON10 \cite{xen7}, see text.}
\label{xcut1}
\end{figure}

\begin{figure}
\begin{center}
\includegraphics[width=15.0cm]{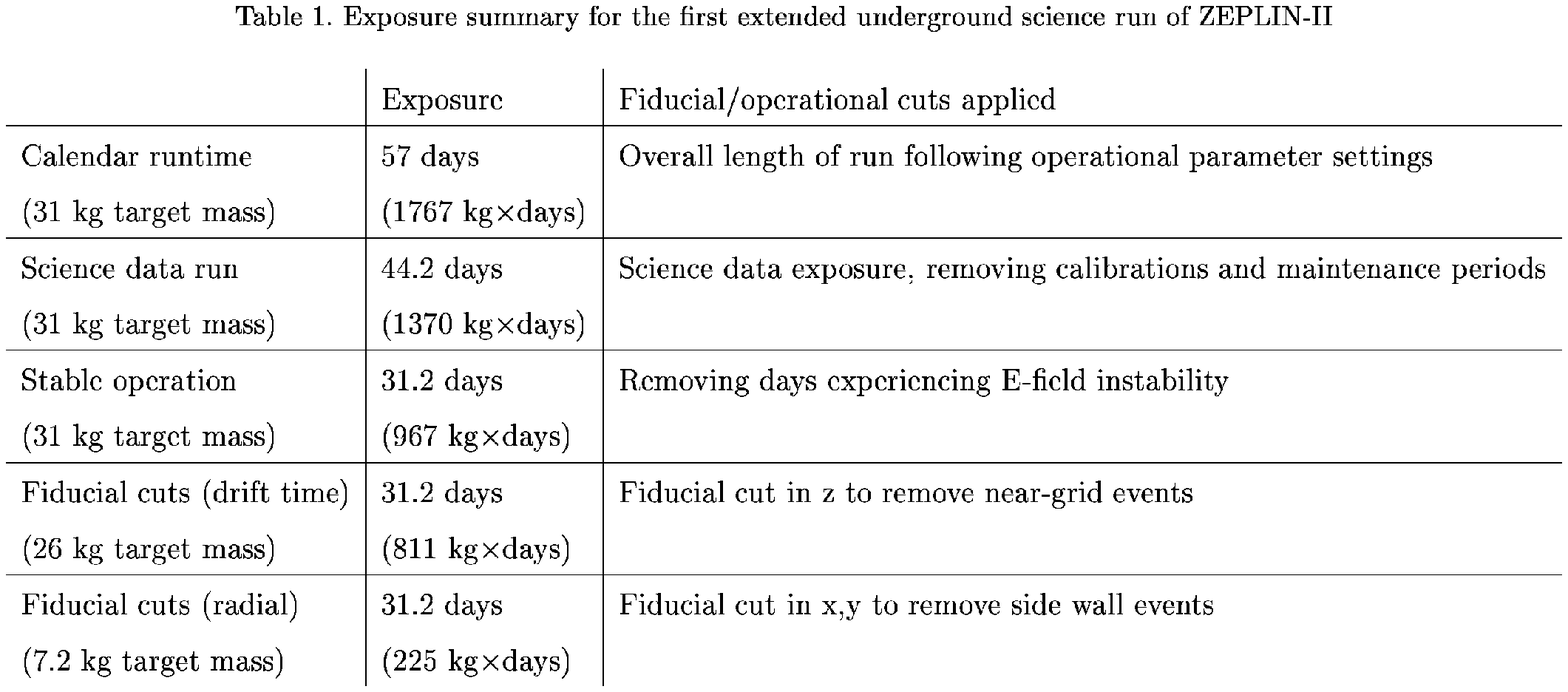}
\end{center}
\caption{List of applied cuts in  ZEPLIN-II \cite{zep3}, see text.}
\label{xcut2}
\end{figure}

\begin{figure}
\begin{center}
\includegraphics[height=9.0cm]{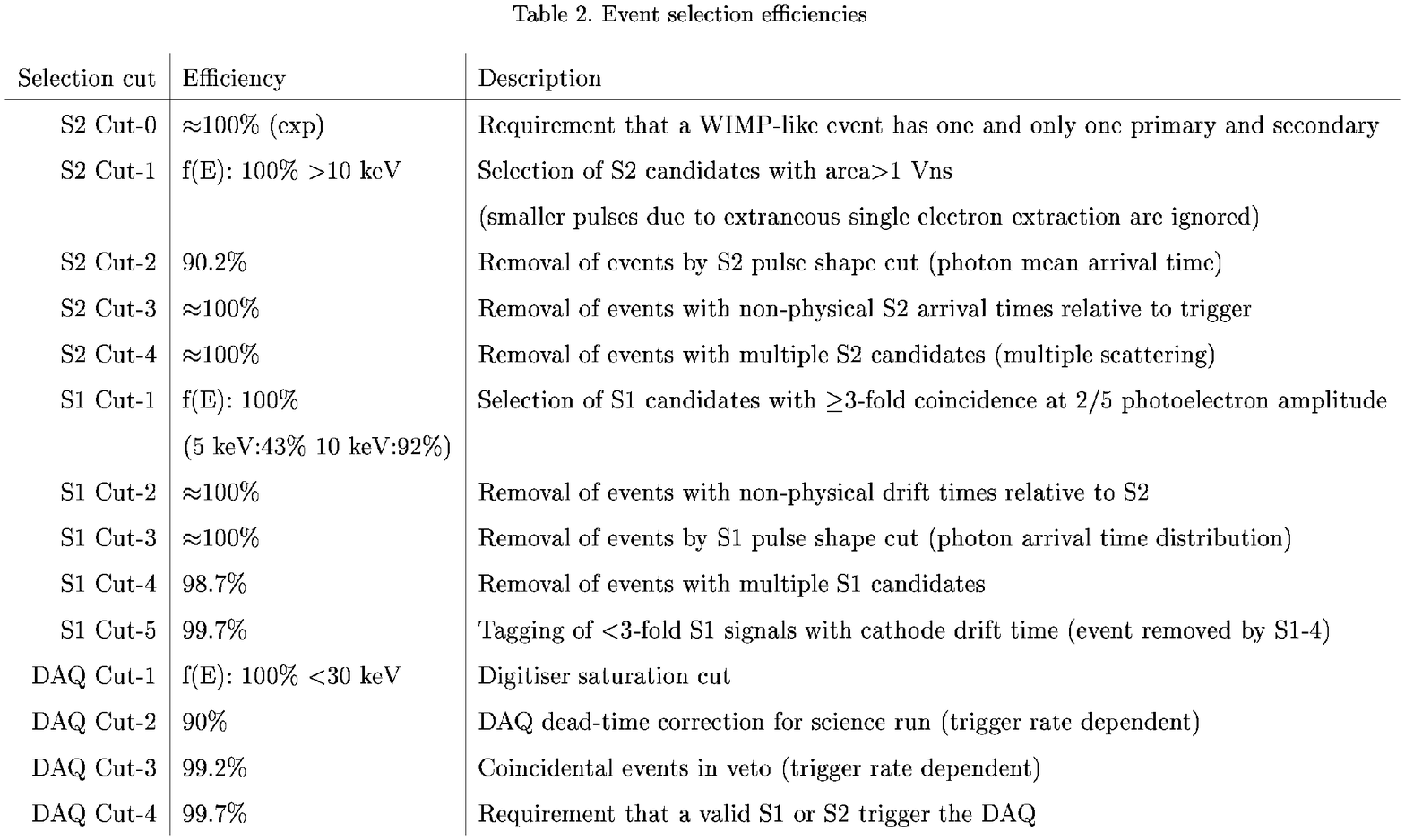}
\end{center}
\caption{List of event selection efficiencies in ZEPLIN-II \cite{zep3}, see text.}
\label{xcut3}
\end{figure}

The list of cuts applied in ZEPLIN-II and  XENON10
are shown in Figs.\ref{xcut1},\ref{xcut2},\ref{xcut3}.

As in XENON10, strong exposure cuts are applied in the first step of analysis in ZEPLIN-II,
in addition to those required by 
$\gamma$ ray calibration periods. ZEPLIN-II also gives the exposure reductions 
due to the fiducial volume cuts;
they excluded periods during which the extraction
field experienced fluctuations in applied voltage \cite{zep3}.

ZEPLIN-II, due to the problem with cooling power, needs to further correct the 
secondary electro-luminescence signal.
In order to keep the time evolution
of S2/S1 distributions stable for nuclear recoil events from the
cathode, corrections are applied to S2: a) for liquid xenon electron lifetime --
{\it the science data and
calibration charge yields were normalized on an event by event basis throughout
the run length};
b) for gas pressure --
{\it The pressure effect on S2 size was measured directly in a
dedicated run and corrected accordingly};
c) for temperature and residual surface charging --
 {\it To account for any residual S2 variability
due to electro-luminescence field variations and liquid surface charging
(due to $<$100\% electron extraction) a final correction was applied.}
These corrections are explicitly mentioned by ZEPLIN-II \cite{zep3}:
{\it Overall the S2/S1 corrections have a maximum variation in log space of  $\pm$10\% from the
mean, excluding the xenon purity correction.}

After these corrections, the data analysis applies a further 14 cuts listed in
 Fig. \ref{xcut3} justified by a number of reasons:
event trigger, DAQ saturation on large secondary pulses etc.
For each cut an energy dependent efficiency is reported as
evaluated from the data and/or simulations; some of the reported efficiencies
(see also Fig. \ref{xcut3}) seem quite optimistic.

\begin{figure}
\begin{center}
\includegraphics[width=13.0cm]{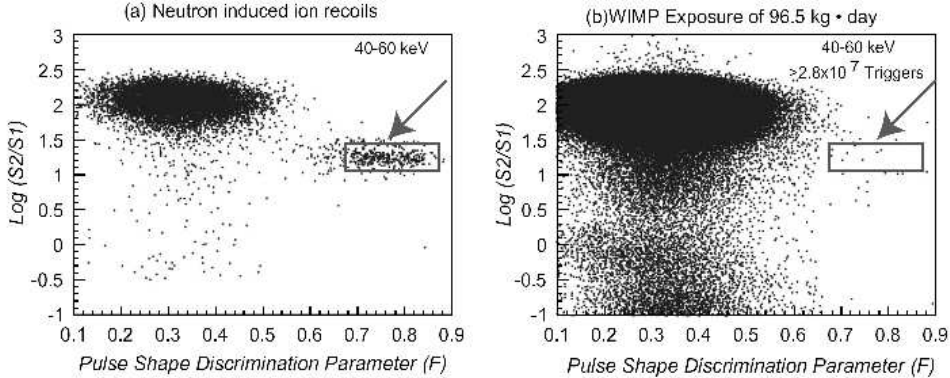}
\end{center}
\caption{WARP: Log(S2/S1) as a function of the pulse shape discrimination parameter
for events obtained with an AmBe source (a) and collected during an exposure of 96.5 kg$\times$day (b).
Figure taken from ref. \cite{arg2}. See text.}
\label{fff}
\end{figure}

The Fig. \ref{fff}, i.e. Fig. 6 of the WARP paper \cite{arg2}, indicates a consistent number of events 
corresponding to an S2/S1 $<$1 in the production data
while there are very few of these events
in the data from AmBe source. The reason for this is not given.\\
Some information
can be derived from XENON10.
In fact in ref. \cite{xen1} the plots produced with sources ($^{137}$Cs
and AmBe) are not shown with a suitable scale to check these events and the production data plot
shows the $\Delta$Log$_{10}$(S2/S1), where they  say that the energy-dependent mean value of
Log$_{10}$(S2/S1) has been subtracted from the electron recoil band
to obtain the $\Delta$Log$_{10}$(S2/S1) for all the events.
In the text on the other hand they explain
that in addition to the leakage of statistical events 
from the electron-recoil band into the nuclear-recoil acceptance 
window they:
{\it observed similar anomalous leakage events 
in the $^{137}$Cs calibration data and unmasked WIMP-search 
data} \cite{xen1}.
The authors identify these events as multiple-scatter
events with 
one scatter in the non-active liquid xenon volume (mostly below 
the cathode)  and a second scatter in the active liquid volume. So they say:
{\it The S2 signal from this type of event is from the interaction 
in the active volume only, while the S1 signal is the sum of 
the two S1's in both the active and the non-active volume} \cite{xen1}. 
The final effect is a resulting smaller S2/S1 value compared to that for a 
single-scatter event, that set these events in 
the WIMP-search window \cite{xen1}. The conclusion is 
that two other cuts are applied in order to remove these events.

These data are not shown in ZEPLIN-II plots, but 
something of similar is mentioned for the calibration data with AmBe
and $^{60}$Co sources performed at a higher acquisition rate than the production runs.
They observe an increased number of 
 random coincidences between real events and events with primary scintillation 
only.  The authors suggest these last events are produced
 from dead regions of the detector where there is a reverse 
drift field. As a consequence: {\it This led to a uniform distribution of events in the 
S2/S1 parameter space, confirmed through study of events with unphysical 
drift times}, i.e. events for which the apparent location is beyond the maximum drift 
distance of the active volume \cite{zep3}. 

The general comments given in this Section also apply to ZEPLIN-III, where the
same kind of rejections and a lot of corrections (not fully quantified in the paper) such as golden event selections,
waveform quality cuts, pulse quality cuts, fiducial volume definition, event quality cuts, etc. 
are applied \cite{zeplas}.

\section{Quenching factors}\label{Qf}

The quenching factor, e.g. the ratio of the response induced by a target recoil nucleus
 to the response induced by an electron of the same kinetic energy, in a given detector
is a necessary experimental quantity for a WIMP direct search involving nuclear recoil.
This quantity is generally measured with a neutron source or by a neutron generator with a specific detector
that reproduces the main features of the underground Dark Matter detector. 
This calibration in situ is avoided in order to prevent  the installation from neutron activations.

Significant differences are found in literature in the measured values even 
for the same nucleus and similar detectors. This may be due to the detector  
or to possible experimental uncertainties.
Some dependence of the recoil/electron response on energy 
has also been reported. The contribution  of the channeling effect
should also be taken into consideration in crystal detectors.

In liquid noble gases, the quenching factor also depends  
on other parameters such as the residual traces of 
impurities acting as quenchers (purity of materials, initial UHV, degassing etc.), thermodynamic
conditions and the presence of an electric field.
The quenching factor in liquid xenon has been measured by many  authors
\cite{BerX,Arn,Aki,Aqf,xen3,xen4,Che}. Less information is
available for liquid argon \cite{arg3,kec} and neon \cite{Mac2}.

In the XENON10 a  200 n/s AmBe source (12 hours of irradiation in-situ) was used to study 
the detector's response to nuclear recoils. The neutron source was shielded by lead to  
reduce the corresponding gamma ray flux \cite{xen1}.
The authors calculated the nuclear recoil equivalent energy 
 as E$_{nr}$ = S1/Ly/Leff$\cdot$(S$_e$/S$_n$). Ly, already quoted in Sect. \ref{Uni} is
 the measured volume-averaged light yield and
 Leff=0.19 is the assumed value for nuclear-recoil scintillation efficiency relative to that of 
122 keV gamma rays in liquid xenon at zero drift field \cite{xen1}. 
S$_e$ and S$_n$ are the scintillation quenching factors, for electron and nuclear recoils,
due to the electric field. S$_e$ and S$_n$ have been given to be 0.54 and
0.93, respectively, at a drift field of 0.73 kV/cm \cite{xen4}.
At the end of the analysis the authors declare that the largest systematics uncertainty is 
due to the limited knowledge of Leff at low 
nuclear-recoil energies. They further say: {\it Our own measurements of this 
quantity [21] did not extend below 10.8 keVr, yielding a 
value of (13.0$\pm$2.4)\% at this energy. More recent measurements
 by Chepel et al. [22] have yielded a value of 
34\% at 5 keVr, with a large error} \cite{xen1}.
Recently a new measurement performed by the XENON10 collaboration on 
Leff\cite{lef}, that has been little studied in the low energy range, shows clear evidence
 of a more general problem: how the uncertainties on the experimental 
 parameters (in this case the Leff value) infer on the final physical conclusions. 
 In fact, as the authors state, the previously claimed\cite{xen1} spin-independent limit, 
  under a single arbitrary\footnote{In fact, a single set of assumptions among the many
  existing possibilities discussed in literature about the astrophysical, nuclear and particle
   Physics aspects is adopted, and neither are experimental and theoretical uncertainties on
    these assumptions generally taken into account. This has already been mentioned in Sect. 1.}
  set of assumptions, is shifted up: 
  for WIMPs of mass 30 GeV/c$^2$  by 24\% and for mass 100 GeV/c$^2$ by 12.5\%,
  just for this reason.

In ZEPLIN-II, nuclear recoil/gamma discrimination has been tested with
$^{60}$Co and Am-Be sources and the authors take
Leff = 0.19, S$_e$ = 0.50, S$_n$ = 0.93 and obtaining  E$_{nr}$ = E$_{ee}$/0.36, 
i.e. as a quenching factor at 1 kV/cm the value Leff$\cdot$(S$_n$/S$_e$) = 0.36.

A similar evaluation is done in ZEPLIN-III, with $^{57}$Co and Am-Be sources.
But when comparing the differential energy spectra for the ``simulation'' curve and the AmBe elastic recoil 
population in S1 electron-equivalent units ($^{57}$Co calibrated S1) there is a significant
mismatch  up to 20 keVee. This disagreement exceeds that expected, as cited by the authors,
 using the calculated efficiencies and threshold effects \cite{zeplas}.
The hypothesis pursued by the authors is non-linearity
in the nuclear recoil energy scale.

\begin{figure}
\begin{center}
\includegraphics[height=8.0cm]{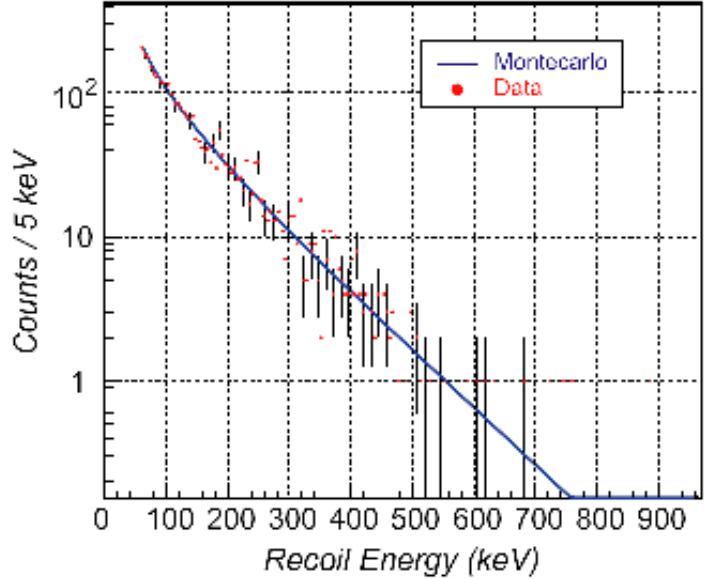}
\end{center}
\caption{WARP: energy spectrum of the recoil-like events collected in the calibration 
with a weak Am-Be neutron source. The data are compared with the Monte Carlo prediction;
the fitted light yield is 1.26 $\pm$ 0.15 phe/keV.
Figure taken from ref. \cite{arg2}.}
\label{ww1}
\end{figure}

The WARP paper does not mention the open questions concerning the quenching factor in Liquid argon
and does not use the word ``quenching factor''. 
The gamma calibrations are only given in a previous paper \cite{arg1} 
at zero field and nothing is said about data on electron
recoils in the energy windows of interest.
They performed calibration with an Am-Be neutron source in situ 
and from the experimental spectrum of $^{40}$Ar
recoil they found a light yield of 1.26$\pm$0.15 phe/keV.
Looking at Fig. \ref{ww1} -- corresponding to Fig. 4 of ref. \cite{arg2} 
from which the light yield for argon recoil is
derived -- it is not clear why, considering the interest
of the low energy region for WIMP detection, the fit in the region below 100 keV is not preferentially shown.
For the quenching factor at 1 kV/cm a value of 0.28 is reported in ref. \cite{arg3} 
for a previous configuration of the 2.3 l detector
viewed by one 8inch PMT. Considering that the recoil light yield is 1.26,
this would correspond to 4.5 phe/keV for the photon light yield, a value that seems too optimistic looking at the energy
resolution and spectra obtained with the $\gamma$ sources and quoted in Sect. \ref{ener}.

A preliminary measurement in liquid argon performed by Micro-CLEAN detector cites a value in the range 0.2--0.3 
for the quenching factor, below 100 keV of recoil energy, without the application of an electric
field \cite{kec}.

In liquid neon the quenching factor has been measured with a portable deute\-rium-deuterium generator
that gives neutrons of 2.8 MeV energy. The authors found a value of 0.26$\pm$0.03 at 387$\pm$11 keV \cite{Mac2} 
at zero field.

\section{Data analysis}\label{Ana}

After the definition of cuts and preliminary rejections let us consider how the different experiments proceed with  
data analysis.

In WARP the published statistics is of 96.5 kg$\times$day, with 2.8 $\cdot$ 10$^7$ triggers that correspond to an integral rate
of 3 $\cdot$10$^5$ counts/day/kg.
After applying the three quoted cuts the authors explain that the pulses surviving:
{\it have been visually scanned. This allows to check the overall pulse shape
of the primary and secondary pulses and to reject mis-reconstructed or noise
events. A total of 1051 events with reconstructed energy $>$40 keV required
visual inspection} \cite{arg2}. Serious doubts arise on 
the methodological approach applied to the analysis. The S2/S1 distribution 
versus energy is not given in the full range for gamma and neutron calibration and for the 
production data, while this distribution would help the reader
understand the effective rejection applied by the analysis. The authors also state
later on in the paper {\it the insufficient sensitivity of rejection
criteria} \cite{arg2}. To be forced to use visual analysis in order to reject
``mis-reconstructed events'' or ``noise'' events suggests the presence of
instrumental problems or an intrinsic problem affecting the application of
an automatic algorithm, so it would also be difficult to estimate the real efficiencies. The authors further comment that 
for 778 events S1 or S2 
or both were not properly calculated by the automatic 
algorithm due to the presence of electronic noise, therefore they recalculated the corresponding values for 
 the S1 and S2 parameters. 
 In 146 events, they were rejected because they were characterised by
 {\it  one of the seven photo-multipliers with a primary signal 
with short rise time and amplitude significantly larger than 
the one detected by the others that, on the contrary, show a 
typical $\gamma$-like behaviour} \cite{arg2}. 
 97 events showed typical behaviour of events from Radon progeny from the walls.
This does not promise well for their future plans towards a 100 kg scale apparatus.

Two energy windows were selected for the WIMP search: 40--60 keV recoil energy 
and 60--130 keV recoil energy.
The only information available
is that 50 photoelectrons correspond to 40 keV of recoil energy.
At the end of analysis five events are found in the window 40-60 keV of recoil energy,
while no other events are found above this energy. In any event they attribute these events 
to spurious origin, neutrons or background
 events and being the maximum recoil energy found 54 keV, the authors evaluate an exclusion plot
for a single set of assumptions above 55 keV, under the hypothesis of zero events.

Further information was added in proof, further statistics collected
of 43 kg$\cdot$day with improved electronics, claiming that {\it when raising the
 threshold to 50 photoelectrons, the measured discrimination for a 70\% nuclear recoil
 acceptance is 3 $\cdot$ 10$^{-7}$} \cite{arg2} and the extrapolated discrimination
 for a 50\% acceptance is better then 1.0$\cdot$10$^{-8}$ . In this case only one nuclear recoil candidate
 was found. It is worth reminding optimistic authors
 that a reliable claim of a 10$^{8}$ rejection power should be supported by 
calibrations and stability monitoring on all aspects involved, with comparable statistics.

XENON10 reported: {\it The analysis was performed blind,
i.e. the events in and near the signal region were not analyzed until the final signal acceptance window and event
cuts were tested and defined} \cite{xen1}. They also have 40
live-days of {\it unmasked} WIMP-search data.
 Based on calibration, and on a period of {\it not-blind} WIMP search 
data, the WIMP search region was defined between 4.5 
and 29.6 keV nuclear recoil energy \cite{xen1}.

ZEPLIN-II describes in more detail the different steps of 
data analysis, the efficiencies and the applied cuts and corrections 
(the energy range of 5-20 keV-- electron equivalent -- was chosen).
Below 5 keV the trigger efficiency is reported to be less than 40\%.
They also use the same expression ``blind''. In fact the authors say that  
{\it To avoid any bias, the selection of cuts and energy range for
the final analysis was based on the results from calibration runs and from unblinded 10\% of data prior
to opening the box with the remaining 90\% of data} \cite{zep3}.
  
However the term ``blind'', often  used in papers written by collaborations using bolometric + 
ionisation/scintillation and liquid noble gases 
techniques is an hyperbole. It is obvious that a correct analysis procedure never decides ``a priori'' the result. 
On the other hand, no information is given on ``unmasked'' data, 
the corresponding rate and energy distributions before any selection and corrections.

In XENON10, an off-line analysis is applied to search the correct S1 signal to be attributed to each trigger.
They require a coinciding signal in at least two PMTs,
stating that under this condition 
 the efficiency of
the S1 signal search algorithm is greater than 99\%, above
a threshold of 4.4 photoelectrons (phe), or 4.5 keV nuclear recoil equivalent energy \cite{xen1}.
The S2 trigger efficiency, with a software
 threshold of 300 phe, they quote it more than 99\% for 4.5 keV
nuclear recoils. 
They also declare that: {\it Basic-quality cuts are used to remove uninteresting events (e.g. multiple scatter and missing S2
events), with a cut acceptance for single-scatter events
close to 99\% } \cite{xen1}.

The detection of nuclear recoil energy is the aim of these different experiments
which apply rejection procedures to the counting rate of the detectors 
in order to identify nuclear recoil populations. 

XENON10 gives the software cut acceptance of nuclear recoils 
and the electron recoil rejection efficiency for 
seven bins in the energy of interest (4.5 -- 26.9) keV nuclear recoil energy,
estimating a number of leakage events, but they say:
{\it However, the uncertainty of the estimated number of leakage 
events for each energy bin in the analysis of the WIMP- 
search data is currently limited by available calibration 
statistics} \cite{xen1}.

Both ZEPLIN-II and XENON10 declare they are able to discriminate
``event by event'', despite large overlapping of populations.
Moreover, too many cuts are applied and even after a most careful study 
of all the data it is still difficult to quantify how many cuts have been made in total.

Only ZEPLIN-II tries to estimate the efficiency for each quoted cut and 
explicitly gives  the exposure surviving each cut. In addition 
each cut can introduce systematics. What about it?
As usual the expression ``event by event'' discrimination, is not properly 
used, which could only be possible in the absence of systematics and when completely separated populations are present,
which is not the case. The rejection procedure of the electromagnetic component
of the counting rate, applied
here and elsewhere, is always a ``statistical'' discrimination on distributions of events and
would  require a much deeper and quantitative investigation than those done up to now on a very 
small scale.
 
In all the experiments at the end, as WARP cited,  
events survive the applied cuts and are ``excluded'' 
on the basis of speculation of being due to WIMP-like processes.

In fact in XENON10 from a total of about 1800 ``final'' events, they observe ten events 
in the WIMP-like 50\% acceptance window, with seven statistical leakage events claimed. But they say: 
{\it Although we treated all 10 events as WIMP candidates in calculating this limit, none of the events are
likely WIMP interactions} \cite{xen1}.

While in ZEPLIN-II 29 events (between 5 keVee and 20 keVee)
survive after the subtraction procedures,
a summed estimate of 28.6$\pm$4.3 $\gamma$-ray and Radon progeny
induced background events are reported. This leads to a 90\%  C.L. upper limit of 10.4 events
candidates as nuclear recoils within the chosen acceptance window \cite{zep3}.

As far as ZEPLIN-III is concerned, an exposure of 847 kg$\cdot$day was published. The authors declare
an energy threshold of 1.7 keVee in S1 (about 10 keV in nuclear recoil energy) 
although the last calibration gamma-line is at 122 keV.  They also declare 
 7 final surviving events in the window of recoils
at the end of the analysis,  which are considered consistent with zero signal\cite{zeplas}.
It not explained why at low WIMP masses, the shown ZEPLIN-III  limit on WIMP-cross section is, for a given
WIMP mass, higher than the XENON10 one
 (considering the two experiments cite approximately the same energy threshold).
 Their conclusion (both ZEPLIN-III and XENON10 results) makes interesting reading:
 {\it However, it is clear that the physics underlying the low-energy performance 
 is poorly understood. 
This is true of  both the response to electron recoils[11]  and to nuclear recoils[12]}.

\section{Miscellany}\label{Mis}

Recent research and development in double phase noble gas detectors
have looked at very low energy applications in Dark Matter direct investigations.
Some further general comments can be done when we look at achieving physical results.

As it is obvious any claimed physical result should be 
 supported and consolidated by: technical papers on the specific detectors and 
their response; long term data taking; publishing sufficiently long exposures;
suitable calibrations in the energy region of interest; studies on stability and so on. In fact experimental
uncertainties can significantly affect any experimental result (i.e. the number of recoil/recoil-like
candidate events) and, therefore, infer on any physical conclusion (i.e. any exclusion plot in whatever scenario).
Let us just take as an example the case of ZEPLIN-I.
In fact, in  ref. \cite{Ben} criticisms were made of the discrimination and background
subtraction procedures applied by ZEPLIN-I, claiming a maximum sensitivity in the spin independent cross section
of 1.1$\cdot$10$^{-6}$ picobarn for a WIMP mass
of 60 GeV under a single set of assumptions \cite{zep1}. 
In particular, the authors of ref. \cite{Ben} have concluded that
the real sensitivity of ZEPLIN-I, in this scenario, would instead be about 10$^{-3}$ 
picobarn. For correctness, 
it is worth noting that a reply by  ZEPLIN-I is in ref. \cite{zep2}; however, it has appeared not well 
convincing due to all the possible
sources of systematics introduced by the lack of stability and reproducibility of the experimental measurements
over time, factors that would be needed to have the claimed sensitivity \cite{zep2}.
Another example is offered by XENON10 that,
despite all the aspects  discussed in previous Sections, has claimed to
be very competitive in some scenario.

On the basis of the arguments discussed in the previous Sections, it is also worth noting that,
despite the discrimination power claimed by the two phase application in noble gases,
 electron and nuclear recoil bands are even more spread and merged
together with respect to those obtained using bolometer/ionisation techniques. 
In addition, non-uniform light collection, non-linear response,
non-particularly competitive energy resolution disfavour their use in low energy physics.  
These features could be partially overcome
by improving some technical aspects, but some may be intrinsic.

In the case of argon the absence of
an odd isotope (useful for investigation into spin dependent WIMP interactions) and
from a technical point of view the need to use wave-length shifter 
for VUV scintillation light collection does not help.
For argon there is still disagreement on results concerning quenching factor 
determination \cite{arg3,kec}.

A comparison of the costs of the different gases can only be made after
an evaluation of the additional costs for Kr-free gas and the reduction
of possible radioactive isotopes as in the case of $^{39}$Ar  for Ar. This is particularly relevant considering the 
fiducial mass is always much lower than the total one, in the present prototypes as well as in future apparata.

In principle, these detectors -- like the cryogenic (bo\-lo\-me\-ter + ionisa\-tion/scin\-til\-la\-tion) ones --
are only potentially sensitive to WIMPs and often only 
to WIMPs belonging to a specific class among the various possible interactions and parameter set.
Beyond this, electromagnetic interaction can play an important role
in a wide range of Dark Matter particles (including scenarios for WIMPs), 
only allowing direct detection when
the rejection procedures of the electromagnetic component of the counting rate are not applied. 

Limitations and methodological bias can also be present in the application of this technology to low energy physics. 
Generally speaking,
the overall aim is to guarantee that, even in the presence of an excess of
 nuclear recoil candidates, these events are due neither to existing side processes nor to
 an instrumental effect (as those discussed in the previous Sections). This issue impoverishes
 not only whatever exclusion plot but also the ``discovery'' power of such experiments in possible future
 configurations with increased masses. Moreover, all the limitations
 on stability, low energy calibrations, rejections, detector response etc. have to be managed 
 for long term data taking.
Finally, if a technological approach allows to define the robustness (in scientific methodology)
of an experimental result,
the last issue interferes ``a priori'' on the reliability of an experimental result \cite{Hud1,Hud2}.

\section{Conclusions}\label{Con}

This paper has analysed and compared the very low energy application
of double phase noble gas detectors.
The main technical aspects of the existing experimental applications have been discussed and
some implications have been outlined. 
The main topics to be addressed in further research and developments have also been presented.

\end{document}